\documentclass[twocolumn,twocolappendix]{aastex63}

\usepackage{amsmath,amstext}
\usepackage[figure,figure*]{hypcap}

\received{receipt date}
\revised{revision date}
\accepted{acceptance date}
\published{published date}
\submitjournal{The Astrophysical Journal}

\shorttitle{Carbonated Silicate Grains}
\shortauthors{Rouill\'e et al.}

\begin{document}

\title{Gas-phase Condensation of Carbonated Silicate Grains}

\author[0000-0002-4016-1461]{Ga\"el Rouill\'e}
\affiliation{Laboratory Astrophysics Group of the Max Planck Institute for Astronomy at the Friedrich Schiller University Jena, Institute of Solid State Physics, Helmholtzweg 3, D-07743 Jena, Germany}
\author{Johannes Schmitt}
\affiliation{Laboratory Astrophysics Group of the Max Planck Institute for Astronomy at the Friedrich Schiller University Jena, Institute of Solid State Physics, Helmholtzweg 3, D-07743 Jena, Germany}
\author{Cornelia J\"ager}
\affiliation{Laboratory Astrophysics Group of the Max Planck Institute for Astronomy at the Friedrich Schiller University Jena, Institute of Solid State Physics, Helmholtzweg 3, D-07743 Jena, Germany}
\author{Thomas Henning}
\affiliation{Max Planck Institute for Astronomy, K{\"o}nigstuhl 17, D-69117 Heidelberg, Germany}

\correspondingauthor{Ga\"el Rouill\'e}
\email{gael.rouille@uni-jena.de}

\begin{abstract}
Reports on the detection of carbonates in planetary nebulae (PNe) and protostars suggested the existence of a mechanism that produce these compounds in stellar winds and outflows. A consecutive laboratory study reported a possible mechanism by observing the non-thermodynamic equilibrium (TE), gas-phase condensation of amorphous silicate grains with amorphous calcium carbonate inclusions. It concluded that water vapor was necessary to the formation of the carbonates. We present a laboratory study with pulsed laser ablation of an MgSi target in O$_2$ and CO$_2$ gases and report, in the absence of water vapor, the non-TE, gas-phase condensation of amorphous carbonated magnesium silicate dust. It consists of amorphous silicate grains with formula MgSiO$_3$ that comprise carbonate groups homogeneously dispersed in their structure. The infrared spectra of the grains show the characteristic bands of amorphous silicates and two bands at $\sim$6.3 and $\sim$7.0~$\mu$m that we assign to the carbonate groups. The silicate bands are not significantly affected at an estimated Si:C ratio of 9:1 to 9:2. Such grains could form in winds and outflows of evolved stars and PNe if C atoms are present during silicate condensation. Additionally, we find that Lyman-$\alpha$ radiation dissociates the carbonate groups at the surface of the carbonated silicate grains and we estimate the corresponding photodissociation cross section of (0.04~$\pm$ 0.02)~$\times$ 10$^{-16}$~cm$^2$. Therefore, photodissociation would limit the formation of carbonate groups on grains in winds and outflows of stars emitting VUV photons and the carbonates observed in protostars have not formed by gas-phase condensation.
\end{abstract}


\section{Introduction}

Carbonates constitute the second most abundant mineral compound after silicates. Largely present on Earth, they have also been detected at the surface of other solid bodies of the Solar system, from planets like Mars \citep[e.g.,][]{Bandfield03,Ehlmann08,Morris10,Horgan20} to asteroids such as Bennu \citep{Kaplan20} and Ryugu \citep{Pilorget22,Nakamura22}, via dwarf-planet Ceres \citep{DeSanctis16,Raponi19}. Accordingly, they have been found in hydrated interplanetary dust particles \citep[e.g.,][]{Sandford86,Tomeoka86,Bradley89,Germani90,Bradley92} and in meteorites \citep{Mittlefehldt94,Scott98,Lee13,Alexander15,Vacher16,Vacher17,Piani18,Ohtaki21}, which are all fragments of the larger solid bodies. All these carbonates are geological materials and their formation proceeds by aqueous alteration of silicates. Thus, the detection of carbonates in a dry place is generally interpreted as evidence of the past presence of liquid water.

Nevertheless, carbonates were spotted in some comets \citep[for a review, see][]{LevasseurRegourd18}, objects that are not thought to contain liquid water in general. Although liquid water may appear in cometary nuclei when specific conditions are fulfilled \citep{Merk06,Prialnik08}, aqueous alteration does not occur ordinarily \citep{Brownlee18} and, consequently, cometary carbonates must have a nebular origin \citep{Wooden08}. Assuming a mechanism exists that led to the presence of carbonates in the solar nebula, it is reasonable that they should appear in other cloud-like environments and that they should figure prominently in the attribution of bands observed in spectra of planetary nebulae (PNe) \citep{Gillett73,Bregman75}. Their discovery in a torus of cold dust in such a nebula was eventually claimed, notably suggesting a formation mechanism in conditions that are inconsistent with aqueous alteration \citep{Kemper02a,Kemper02b}. Reports of carbonates in protostars \citep{Ceccarelli02,Chiavassa05} and around a young star \citep{Lisse06,Lisse07a} followed. Recently, \citet{Bowey22} proposed that carbonates contribute to the absorption spectrum of the dust torus in Sakurai's Object (V4334~Sgr) and \citet{Bowey23} estimated the spectral contribution of crystalline carbonates to extinction by molecular clouds and in young stellar objects.

Astronomical observations prompted \citet{Toppani05} to test the formation of carbonates in the winds of evolved stars and outflows of protostars. They synthesized amorphous grains by gas-phase condensation of silicate vapors produced with laser shots at glassy targets of mixed oxides in a low-pressure gas. When this gas contained both carbon dioxide (CO$_2$) and water vapor (H$_2$O), the IR spectra of the grains featured carbonate bands and the grains contained carbonate domains. To explain the presence of carbonate inclusions, \citet{Toppani05} proposed a two-step chemistry in conditions off thermodynamic equilibrium, specifically, the hydration of silicate molecular clusters followed with carbonation. \citet{Rietmeijer08} argued that a two-step mechanism might not be efficient and proposed that carbonates were formed through reaction of amorphous silicate grains with CO$_2$ and, possibly, carbon monoxide (CO) molecules. Both hypotheses assumed adequate nebular conditions.

We present a study on the gas-phase condensation of amorphous grains in conditions relevant to outflows of evolved star. The grains consist of a silicate matrix containing carbonate groups and we test the importance of water in their formation. We then examine the relevance of the grains produced by gas-phase condensation to explaining the detection of carbonates in PNe and protostars. Finally, bearing in mind the still unexplained depletion of interstellar gas-phase O atoms \citep{Jenkins09,Whittet10,Jenkins19}, we discuss the relevance of carbonates to assessing the amount of oxygen locked in cosmic dust.

\section{Experimental}\label{sec:exp}

The apparatus for the gas-phase synthesis of refractory grains is described in detail in a study on the preparation of interstellar silicate analogs \citep{Sabri14}. Presently, a metal target with composition MgSi (Evochem Advanced Materials GmbH, 99.5\% purity) is vaporized with laser shots (532~nm wavelength, shots of 10~ns duration, 10~shots per second, 110~mJ per shot) while under a continuously refreshed, low-pressure ($\sim$6~Torr) atmosphere of O$_2$ (Air Liquide, 99.9995\% purity), CO$_2$ (Westfalen, 99.995\% purity), or a mixture thereof. Focused at the surface of the rotating target, the nanosecond-long laser shots generate each a localized plasma that expands extremely rapidly into the gas phase. Collision of the plasma with the ambient gas forms an expansion front where heated and compressed plasma and ambient gas interact, causing at least partial dissociation of a fraction of the molecules that compose the latter. As the plasma expands, the temperature in the front decreases rapidly relative to the density, leading to supersaturation conditions and enabling the condensation of gas-phase components into grains \citep[][and references therein]{Kautz22}. The fast quenching rate of up to 10\,000~K~s$^{-1}$ leads to solids with an amorphous structure.

Grains are extracted from the chamber and formed into a beam by means of two-fold differential pumping, first through a nozzle, then through a skimmer. The beam is directed through a gate valve toward a fourth chamber where the solid particles are deposited onto a substrate mounted on the cold head of a compressed-helium, closed-cycle cryocooler (Advanced Research Systems, Inc. ARS-4H and DE-204SL) that allows us to cool the substrate and the deposited grains down to $\sim$7~K. The cryocooler can rotate to make the substrate face various ports dedicated to in situ spectroscopy or irradiation of deposits. We carry out mid-IR absorption spectroscopy by way of transmission measurements through opposite ports equipped with Tl(Br,I) windows. These ports are connected to an evacuated FTIR spectrometer (Bruker VERTEX 80v) on one side and the evacuated chamber of its external detector on the opposite side. We use 2~mm thick KBr substrates (Korth Kristalle GmbH) for mid-IR measurements.

For studying the effect of vacuum ultraviolet (VUV) irradiation on the grains, a port of the chamber is fitted with a 3~mm thick MgF$_2$ window and a microwave-discharge H$_2$-flow lamp. The glass of the lamp has an F-type shape \citep{Ligterink15} and is made of quartz. We make H$_2$ gas (Air Liquide, 99.9\% purity) flow through the lamp by applying a constant pressure of 1~mbar at one end and evacuating the gas with a vacuum pump at the other. We estimate the pressure to be $\sim$0.5~mbar in the lamp. A 2450~MHz microwave generator (SAIREM GMP 03 KSM B) feeds an Evenson cavity (Opthos) attached to the lamp and excites the gas with a forward power of $\sim$90~W and a typical reflected power of $\sim$2~W. The distance from the center of the cavity to the sample is 14~cm, from the MgF$_2$ window to the sample, 31~mm. Applying the method of \citet{Fulvio14}, we evaluate the Lyman-$\alpha$ photon irradiance at sample to be (145~$\pm$ 50)~$\times$ 10$^{12}$~cm$^{-2}$ s$^{-1}$. It corresponds to a photoelectric current of 0.9~$\pm$ 0.3~$\mu$A emitted by a golden surface 14~mm in diameter. It takes into account the difference of the window material of the photoelectric measurement chamber (1.5~mm LiF instead of 3~mm MgF$_2$) and a proportion of Lyman-$\alpha$ photons of 87\% measured using a sapphire filter. The irradiance is consistent with data obtained by \citet{Ligterink15} for a distance of 31~cm between window and sample.

Emission spectra are measured over the wavelength range 350--1150~nm during laser ablation. Because the spectrometer (Ocean Optics Inc. QE65000) does not allow us to carry out time-gated measurements, photons are counted during time intervals that see several ablation events and we use a high-reflectance dielectric mirror (Laseroptik GmbH, 532~nm central wavelength, 0$^{\circ}$ angle of incidence) as a filter to attenuate the laser light as strongly as possible. To possibly reveal weak emission lines while avoiding saturation of the sensor, photons are counted and accumulated for 10 consecutive ablation events.

We carry out high-resolution transmission electron microscopy (HRTEM) and energy-dispersive X-ray (EDX) spectroscopy of the grains ex situ by transferring grains from a deposit to a substrate for HRTEM, namely, a lacey carbon film supported by a copper grid. The microscope is a high-resolution scanning transmission electron microscope (STEM; JEOL GmbH JEM-ARM200F NEOARM) operated with an acceleration voltage of 200~kV and it is equipped with an EDX silicon drift detector that enables elemental analysis in scanning mode. For structure examination and elemental analysis, we select exclusively particles that stand above holes in the carbon film. Complete quantitative elemental analysis is not given owing to the large uncertainty on the scaling of values for C and O atoms.

\section{Calculations}\label{sec:calc}

Quantum chemical calculations assisted us in characterizing nanometer-sized grains of amorphous carbonated silicates, in particular their carbon contents. We borrowed the theoretical structures of a nanosilicate with enstatite pyroxene (MgSiO$_3$) composition from \citet{Escatllar19}. In this structure of formula (MgSiO$_3$)$_{10}$, we replaced one of the Si atoms with a C atom, optimized the geometry of the new grain, and computed its harmonic vibrational frequencies. We repeated the operation for every Si atom to obtain frequencies for a variety of positions of the C atom, and thus increased the probability to come across a coincidence with the measured spectra assuming the synthesized grains were amorphous carbonated silicates.

The theoretical properties were computed with the Gaussian software \citep{Gaussian16}. We chose a model chemistry that used density functional theory and consisted of combining the B3LYP functional \citep{Becke88,Becke93,Lee88,Stephens94} with the 6-31g(d) basis set \citep{Hariharan73,Hariharan74}. When mentioned, theoretical atomic charges are derived from computed atomic polar tensors \citep[APT charges,][]{Cioslowski89}.

\section{Results}

\subsection{Synthesis}

\begin{figure}
\epsscale{1.1}
\plotone{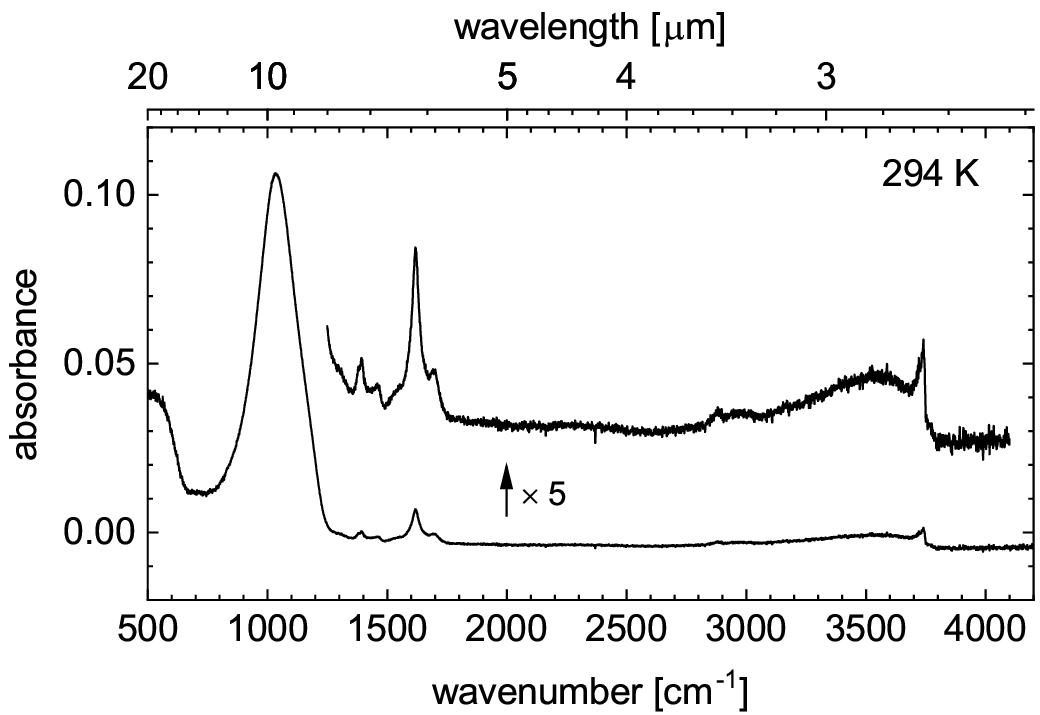}
\caption{Absorption spectrum of grains obtained from laser vaporization of an MgSi target in O$_2$ gas. Measurement carried out in situ at 294~K immediately after deposition and reported without baseline correction. The 1250--4100~cm$^{-1}$ range is reproduced vertically shifted and expanded for a better display of details.\label{fig:MIR-MgSi-O2}}
\end{figure}

Figure~\ref{fig:MIR-MgSi-O2} shows the IR spectrum of the material produced by laser ablation of an MgSi target in O$_2$ gas. It features, at $\sim$10 and $\sim$20~$\mu$m wavelength, the characteristic broad absorption bands of amorphous silicate grains \citep{Sabri14}. Because the composition of the target drives that of the grains formed by condensation \citep{Fabian00,Sabri14}, we identify the sample as a deposit of amorphous silicate grains with the composition of enstatite, MgSiO$_3$, and discuss its IR spectrum further in Section~\ref{sec:MIR-Sil}.

\begin{figure}
\epsscale{1.1}
\plotone{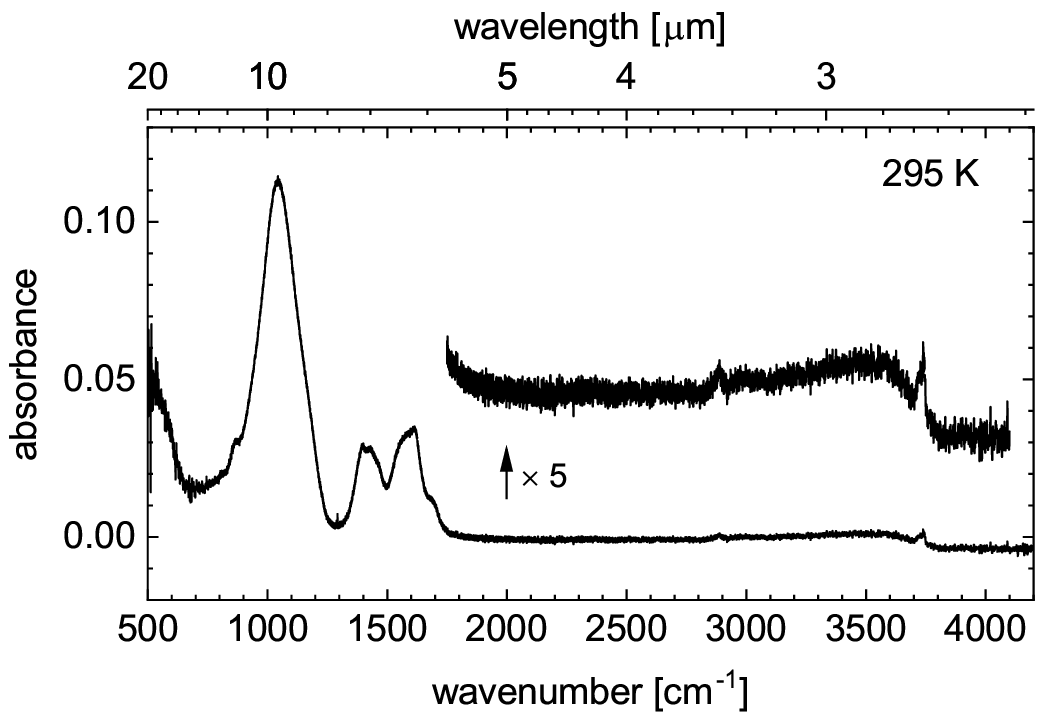}
\caption{Absorption spectra of grains obtained from laser vaporization of an MgSi target in CO$_2$ gas. Measurement carried out in situ at room temperature (295~K) immediately after deposition. The spectrum is reported without baseline correction. The 1750--4100~cm$^{-1}$ range is reproduced vertically shifted and expanded for a better display of details.\label{fig:MIR-MgSi-CO2}}
\end{figure}

Figure~\ref{fig:MIR-MgSi-CO2} demonstrates that, when replacing the O$_2$ atmosphere with CO$_2$ gas, we still obtain grains and the main features in their IR spectrum are the silicate bands at $\sim$10 and $\sim$20~$\mu$m. The spectrum also shows two new absorptions at $\sim$6.3 and $\sim$7.0~$\mu$m (at $\sim$1580 and $\sim$1420~cm$^{-1}$, respectively, in terms of wavenumber), both of medium strength compared to the strongest silicate band. We attribute them to the asymmetric stretching of CO bonds in carbonate groups and, given a similarly separated doublet in the spectrum of an Mg-rich carbonated silicate melt \citep{Brooker01b}, we postulate that the sample consists of carbonated silicate grains. Section~\ref{sec:MIR-Mix} proposes a detailed analysis of the spectrum. We remark that the strength of the new absorption bands depends on the C:O ratio in the ambient gas atmosphere as demonstrated in Appendix~\ref{apx:gas-comp} with the use of O$_2$:CO$_2$ mixtures.

\begin{figure}
\epsscale{1.1}
\plotone{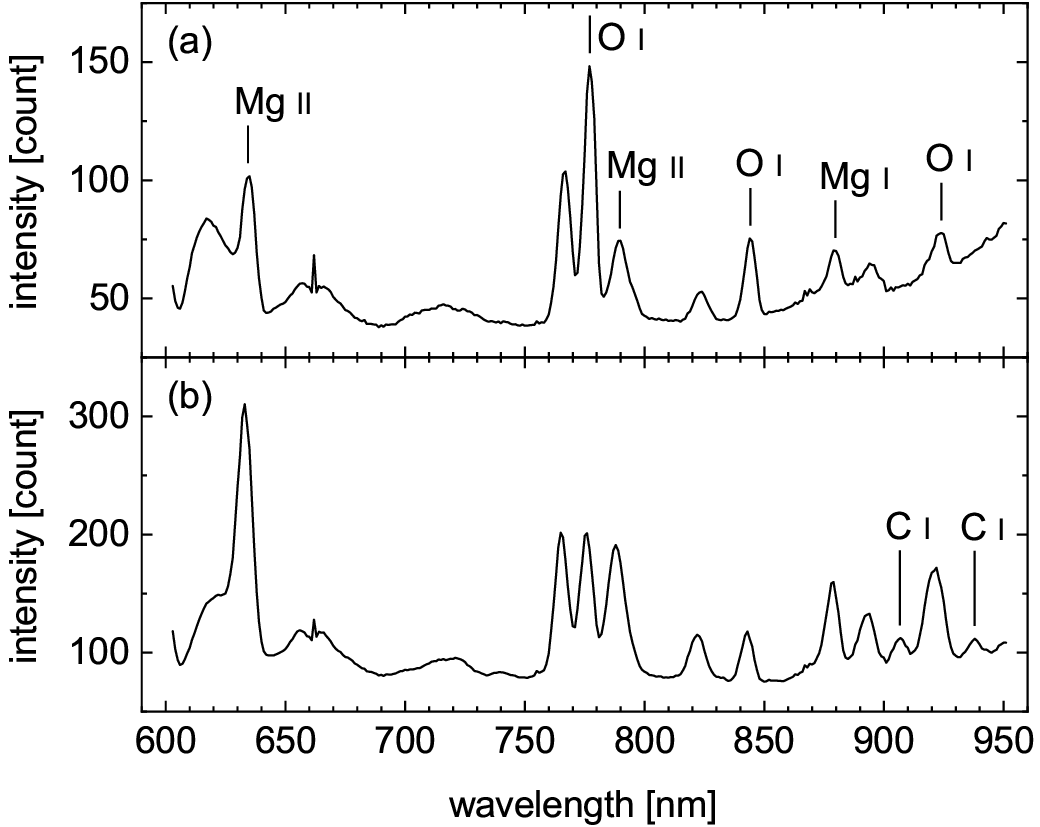}
\caption{Emission spectra obtained during laser vaporization of an MgSi target in atmospheres of O$_2$ (a) and CO$_2$ (b). Labels indicate the assignment of identified lines.\label{fig:E-MgSi-O2-CO2}}
\end{figure}

The fact that silicate grains are formed from an oxygen-free target when the ambient gas is O$_2$ or CO$_2$ demonstrates that reactive laser ablation occurs \citep{Johnston92}. Molecules of the ambient gas diffuse into the front of the expanding plasma and are consequently ionized and dissociated to an extent that depends on the power density of the laser light at the target and on the pressure of the ambient gas. Condensation combines the products of this process with the elements of the plasma that were initially components of the target. Emission spectra measured during experiments, such as those displayed in Figure~\ref{fig:E-MgSi-O2-CO2}, feature \ion{O}{1} lines when the ambient gas is O$_2$, \ion{O}{1} and \ion{C}{1} lines when it is CO$_2$, and thus confirm the dissociation of molecules of the ambient gas. Some degree of ionization is demonstrated by the \ion{Mg}{2} lines. Considering that silicates are formed by condensation, we propose that amorphous carbonated silicates and possibly carbonates are produced in the same way. Consequently, a mechanism that does not require water in any form is available for the synthesis of these materials.

\subsection{Mid-infrared Spectroscopy}\label{sec:MIR}

\subsubsection{Silicate Grains}\label{sec:MIR-Sil}

Figure~\ref{fig:MIR-MgSi-O2} shows the spectrum of grains at room temperature obtained by using O$_2$ as the reactive gas during laser ablation of an MgSi target. The spectrum is dominated by the bands of amorphous silicate matter \citep{Sabri14} that rise at $\sim$10 and $\sim$20~$\mu$m, the strongest one peaking at 9.65~$\mu$m or 1036~cm$^{-1}$ in terms of wavenumber. The values are close to those observed for amorphous silicate grains with enstatite formula, MgSiO$_3$ \citep[see constants and spectra in, e.g.,][]{Scott96,Fabian00}. Elemental analysis with EDX, not shown, verifies this composition.

We assign the minor absorptions near 2900~cm$^{-1}$ (3.4~$\mu$m wavelength) to the CH-stretching modes of aliphatic hydrocarbon structures. These structures may be products of the grain condensation process, their components coming from the dissociation of background-gas molecules contaminating the O$_2$ atmosphere. They may also be substances of the background gas adsorbed physically or chemically by the silicate grains after their deposition. In either case, tests showed that KBr substrates do not adsorb aliphatic substances in measurable quantities during experiments at room temperature, only grain deposits do.

The broad, low feature with a maximum at $\sim$3550~cm$^{-1}$ corresponds to the stretching of O$\sbond$H bonds in water molecules adsobed by the grains and also in silanol (Si$\sbond$OH) groups caught in bound vicinal configuration \citep[e.g.,][with silicas]{McDonald58,Burneau90}. As to the sharp absorption at 3740~cm$^{-1}$, it arises from O$\sbond$H stretching in free silanol groups, the tail of this absorption toward lower wavenumbers coming from terminal vicinal silanol groups \citep{McDonald58,Burneau90}.

Minor absorption bands are visible between 1300 and 1800~cm$^{-1}$ where the bending vibration of adsorbed water molecules is active as well as deformation modes of aliphatic hydrocarbon groups. Appendix~\ref{apx:MIR-fits} presents an analysis of these bands with fitted Gaussian profiles. Given the strength of the OH-stretching signal near 3500~cm$^{-1}$, to which water is not the only contributor, we expect the bending vibration of water to cause an absorption weaker than the bands seen at $\sim$1620 and $\sim$1700~cm$^{-1}$, hence it is either blended with them or too weak to affect the spectrum. Similarly, seeing the comparatively weak aliphatic CH-stretching bands visible near 2900~cm$^{-1}$, we do not assign the bands between 1350 and 1500~cm$^{-1}$ to deformation modes of aliphatic hydrocarbon groups. Consequently, and in view of observations made during VUV irradiation experiments (see Section~\ref{sec:irrad}), we assign the minor absorption features between 1300 and 1800~cm$^{-1}$ to the asymmetric stretching of CO bonds in carbonate groups with various binding configurations \citep[see][for instance, and references therein, for information on vibrational frequencies of carbonate groups at the surfaces of various materials]{Taifan16}. We do not see bands that could correspond to other vibrational modes of carbonate groups, possibly because they are weaker. We attribute the formation of the groups to components of the background gas in the laser ablation chamber of the apparatus. As the primary base pressure is $\sim$0.05~mbar, the phenomenon indicates that a low concentration of a given substance in the ambient gas during laser ablation can affect measurably the composition of the grains being produced. We do not exclude the possibility that the groups form after the condensation of the grains, even as late as at the time of their deposition. In that case the groups would be bonded to the surface of the grains.

\subsubsection{Carbonated Silicate Grains}\label{sec:MIR-Mix}

Figure~\ref{fig:MIR-MgSi-CO2} shows the spectrum of a deposit at room temperature obtained by using an MgSi target and CO$_2$ as the reactive gas instead of O$_2$. All absorption features observed above in the spectrum of the amorphous silicate deposit are present, including the minor ones. The peak position of the strongest silicate absorption, however, has shifted from 1036 to 1044~cm$^{-1}$. In the absence of any perceptible contribution by a new band superimposed on the silicate absorption peak, we attribute the shift to a different composition and structure of the grains, that is, the insertion of C atoms and extra O atoms in the MgSiO$_3$ structure. Actually, lowering the degree of polymerization of the silicate groups, i.e., the average number of bridging O atoms, shift the 10~$\mu$m band toward lower wavenumbers \citep[e.g.,][]{Mysen82,McMillan84,Jaeger03b,Jaeger03a}. Elemental analysis with EDX verifies the composition of the silicate material and the distribution of C atoms in the grains (Section~\ref{sec:EMEA-Mix}).

New major spectral features consist of two broad absorptions that rise under the minor bands of the 1300--1800~cm$^{-1}$ range that we have already mentioned. We attribute them to CO stretching vibrations in carbonate groups and rule out structures that contain hydrocarbon groups or even OH groups, for instance, carboxyl and bicarbonate groups. Indeed, the rise of the new absorptions is not correlated with an increase of absorption at CH- and OH-stretching frequencies, near 2900~cm$^{-1}$ and in the 3000--3700~cm$^{-1}$ range, respectively (see also Appendix~\ref{apx:gas-comp}). The vibration of the carbonate ion relevant to the rise of the two bands is the doubly degenerate asymmetric CO stretching mode $\nu_3$. Its present manifestation as a doublet indicates that the degeneracy of the mode is lifted and hence that the binding of the carbonate groups to the structure of the condensate is not symmetric.

Taking this assignment into account, the 8~cm$^{-1}$ blue shift of the strongest peak to 1044~cm$^{-1}$, mentioned above, could result from the rise of a new band. Although the symmetric CO stretching mode of carbonate groups, $\nu_1$, is not IR active in the free CO$_3^{2-}$ ion, the conditions that lift the degeneracy of the $\nu_3$ mode make the $\nu_1$ mode IR active. The strength of the corresponding band may be too low to affect the position of the 10~$\mu$m peak, however. As we do not see signs that the band is present, we are left with a shift induced by a change in the degree of polymerization of the silicate groups following the insertion of carbonate groups in the silicate structure \citep[see network carbonates in][and references therein]{Brooker01b}. Elemental analysis and observations during VUV irradiation of the sample are in agreement with this conclusion (Sections~\ref{sec:EMEA-Mix} and \ref{sec:irrad-Mix}).

The appearance of a minor band at 862~cm$^{-1}$ coincides with that of the bands attributed to carbonate groups and we could assign it tentatively to $\nu_2$, the out-of-plane bending mode (inversion) of these groups. \citet{Ciaravella16} actually attributed to carbonates what appears to be the same band observed in the spectrum of a sol-gel silicate. Alternatively, it could be a silicate feature induced by the presence of carbonate groups in the material. Already we have attributed the shift of the 10~$\mu$m band compared to the pure silicate sample to a change in the degree of polymerization of the silicate groups. A relative increase of the number of SiO$_4$ tetrahedra with four non-bridging O atoms may give rise to the band at 862~cm$^{-1}$. Again, observations during VUV irradiation of the sample are in agreement with this conclusion (Section~\ref{sec:irrad-Mix}).

We note that silanol groups are again observed, though not in greater amount than in the pure silicate deposit. Interestingly, the composition of the grains formed by reactive ablation in CO$_2$ gas does not prevent the existence of free silanol groups that are revealed by the narrow band at 3739~cm$^{-1}$. We conclude that hydrogen-bonding does not occur between the carbonate and silanol groups and infer that the carbonate groups are polydentate rather than mono- or bidentate.

\subsection{Electron Microscopy and Elemental Analysis}\label{sec:EMEA-Mix}

\begin{figure}
\epsscale{1.1}
\plotone{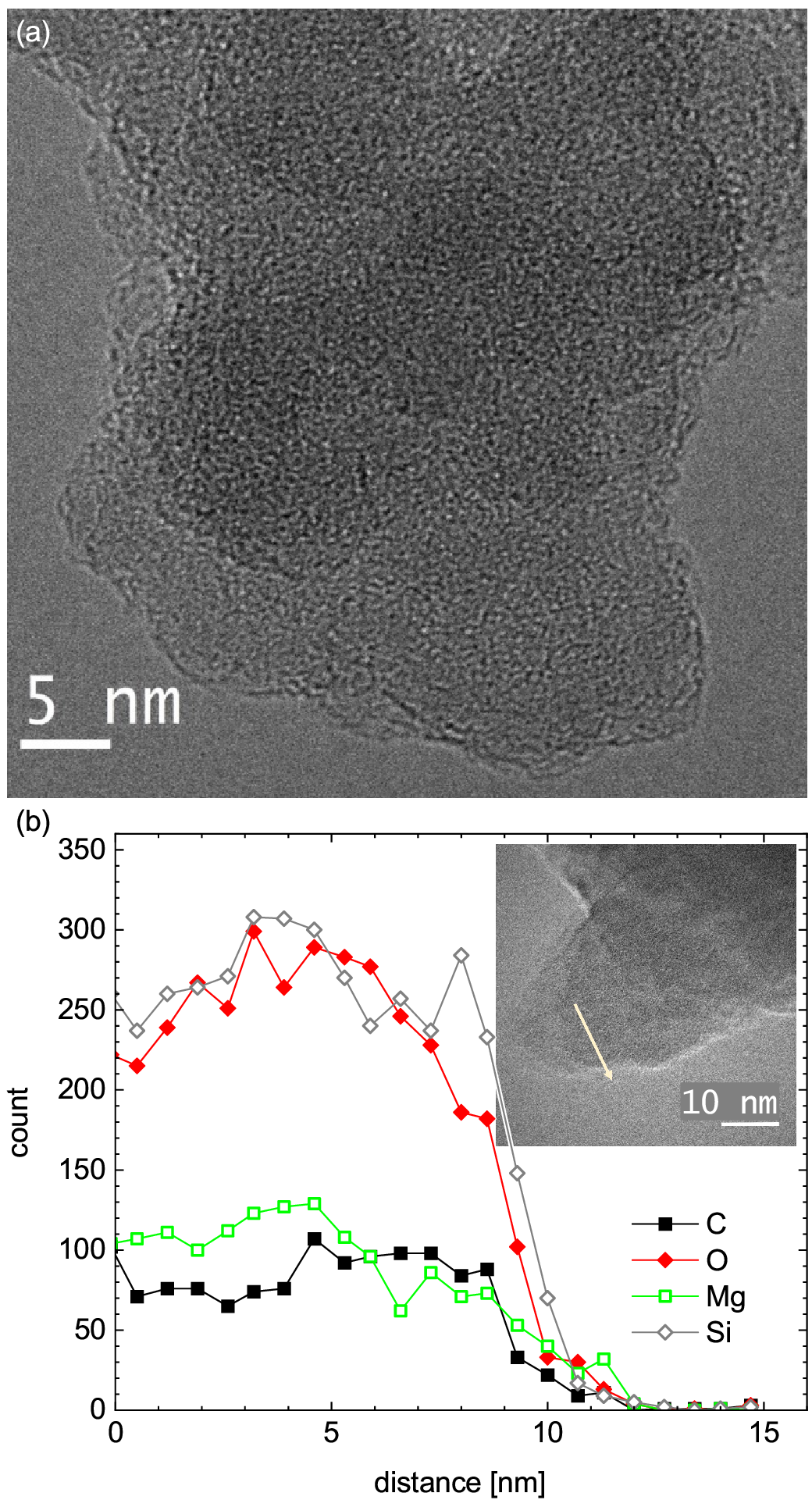}
\caption{Microscopy of carbonated silicate grains. Image taken at room temperature with HRTEM revealing the amorphous internal structure of the nanometer-sized grains in an agglomerate (a). Elemental analysis of grains along a line with EDX spectroscopy (b). An arrow drawn in the inset image represents the line scanned to obtain the measurements. The vertical axis gives the  K-line emission intensity as photon count for each element, not the number of atoms. The measurements for C, O, Mg, and Si are represented with black full squares, red full diamonds, green empty squares, and gray empty diamonds, respectively, connected with solid lines of the corresponding color. In each panel, a horizontal bar and its length in units of nanometer indicate the scale.\label{fig:TEM}}
\end{figure}

Figure~\ref{fig:TEM} shows an HRTEM image of carbonated silicate grains transferred from a deposit and a graph representing the distribution of elements through grains as measured with EDX spectroscopy. The HRTEM image (Figure~\ref{fig:TEM}a) reveals that the deposit consists of agglomerated nanometer-sized grains or nanoparticles and confirms that their internal structure is amorphous. Furthermore, the elemental analysis of grains along a line (Figure~\ref{fig:TEM}b) demonstrates that the distribution of the elements, including carbon, is homogeneous. Examination of several grains gave consistently similar results. Thus, we conclude that the carbonate groups are distributed throughout the silicate network and that this carbonated silicate is the product of gas-phase condensation. It also verifies that the silicate composition is MgSiO$_3$ according to ion-count calibration with internal standards and also with our own standard materials such as glassy MgSiO$_3$ and forsterite (Mg$_2$SiO$_4$).

\subsection{Irradiation}\label{sec:irrad}

\subsubsection{Silicate Grains}\label{sec:irrad-Sil}

\begin{figure}
\epsscale{1.1}
\plotone{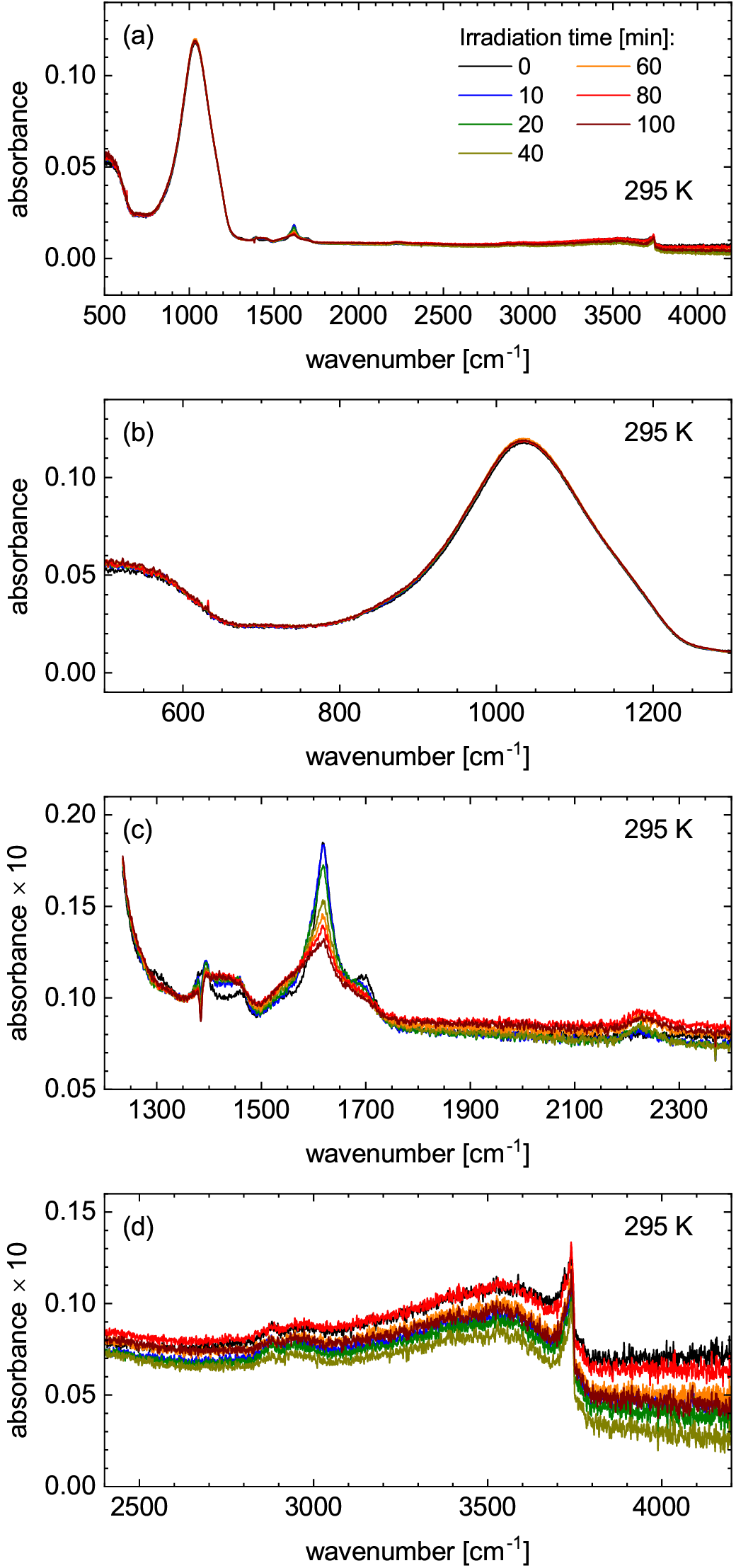}
\caption{Effect of VUV irradiation on the infrared spectrum of grains produced by laser vaporization of an MgSi target in an O$_2$ atmosphere, with substrate holder at 295~K during irradiation and during spectroscopy. Wide view of the spectra (a), focus on the silicate bands (b), on minor absorptions (c), and on bands corresponding to CH- and OH-stretching vibrations (d). The raw absorbance spectra, measured consecutively, were offset vertically by an arbitrary amount determined from the average absorbance between 1345 and 1355~cm$^{-1}$ so as to superimpose their baselines.\label{fig:VUV-FTIR-RT-O}}
\end{figure}

Figure~\ref{fig:VUV-FTIR-RT-O} illustrates the effect of VUV irradiation on the sample of amorphous silicate grains examined in Section~\ref{sec:MIR-Sil}. We carried out the irradiation in situ shortly after producing the deposit. The 10~$\mu$m silicate peak is mostly stable during irradiation.
Nevertheless, while its height varies by no more than $\pm$1\% without any clear trend, its width increases slightly on the lower wavenumber (longer wavelength) side as a function of irradiation time. As to the 20~$\mu$m band, absorbance increases by $\sim$10\% at 500~cm$^{-1}$, indicating that the band gains strength (Figure~\ref{fig:VUV-FTIR-RT-O}b). The changes suggest a very slight rearrangement of the silicate structure.

The minor absorption bands observed between 1300 and 1800~cm$^{-1}$, attributed in Section~\ref{sec:MIR-Sil} to variously bonded CO$_3$ groups, are strongly attenuated. Absorption features outside this wavenumber interval do not show clear attenuation. A fine analysis of the broad absorption caused by O$\sbond$H stretching in silanol groups and adsorbed water molecules is impracticable because the baseline is irregular and varies from a spectrum to the next (Figure~\ref{fig:VUV-FTIR-RT-O}d). Nevertheless, a comparison of the spectra measured before irradiation and after 100~min irradiation does not reveal any visible change in the broad feature except for a minor difference at 3700~cm$^{-1}$, at the foot of the narrow band caused by the OH-stretching mode of free silanol groups. Thus, we confirm that the bands in the 1300--1800~cm$^{-1}$ range, attenuated by irradiation, are not related to OH groups and are the mark of carbonate groups. Moreover we find a first hint that VUV irradiation can destroy these carbonate groups.

The resilience of the silanol groups may seem surprising given that Lyman-$\alpha$ radiation can actually decompose them. The rate of the process, however, is slow in comparison with the duration of our experiment \citep{Rajappan11}. Moreover, regeneration of silanol groups can occur through reaction between background H$_2$O molecules that are present in the high-vacuum chamber (base pressure of $\sim$10$^{-7}$--10$^{-6}$~mbar at room temperature) and Si atoms with dangling bonds \citep{Lee10}. Additionally, absorption of VUV light by the grains partially shields the groups.

Upon irradiation arises a weak, broad absorption centered at $\sim$2240~cm$^{-1}$ (Figure~\ref{fig:VUV-FTIR-RT-O}c). We assign it to the stretching of Si$\sbond$H bonds that cause absorption at 2246~cm$^{-1}$ when present at the surface of amorphous grains with enstatite (MgSiO$_3$) composition \citep{Blanco99}. We hypothesize that SiH groups issue from the dissociative adsorption of water molecules at Si atoms with dangling bonds created by VUV irradiation. The mechanism produces silanol (Si$\sbond$OH) groups at the same time and contributes to maintaining their number.

\subsubsection{Carbonated Silicate Grains}\label{sec:irrad-Mix}

\begin{figure}
\epsscale{1.1}
\plotone{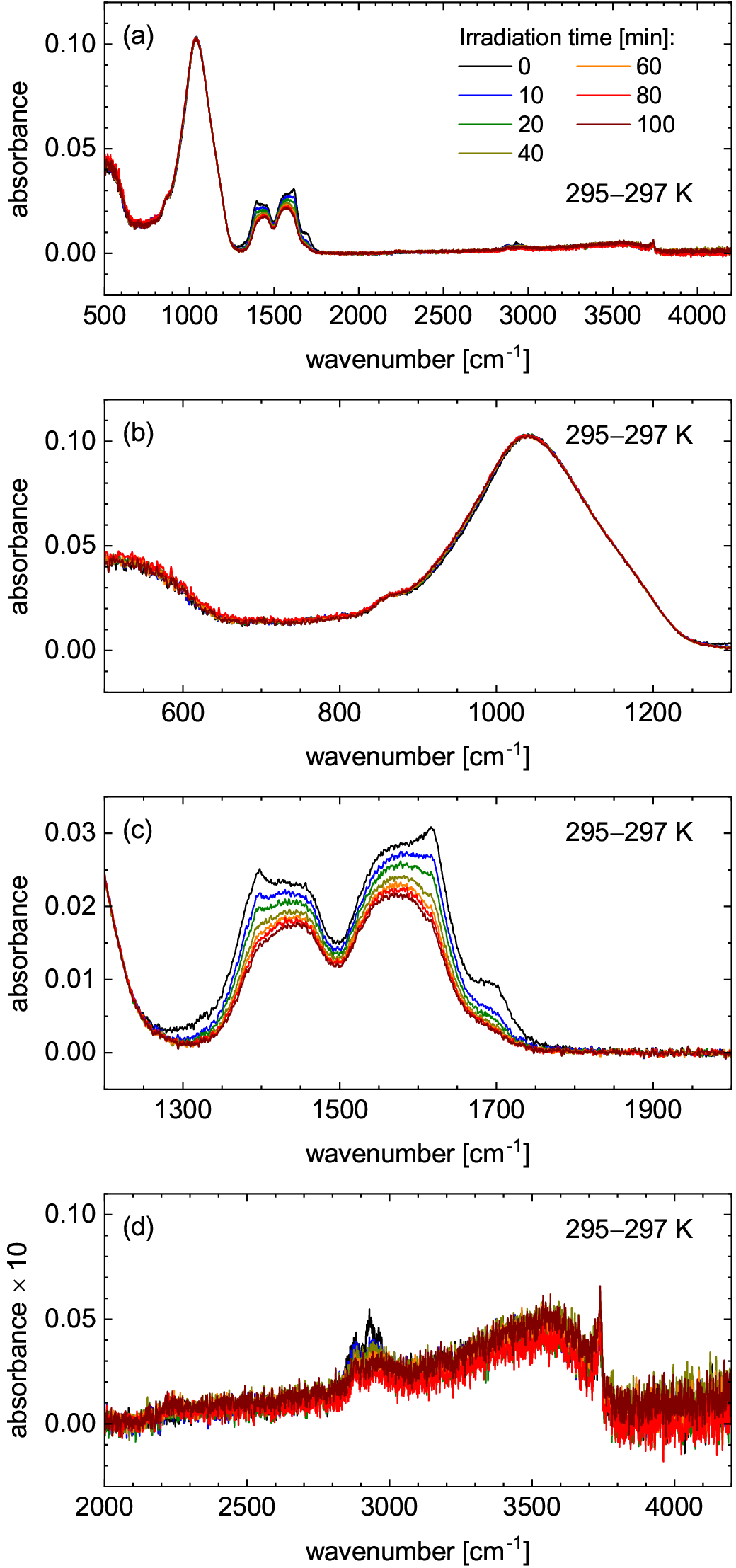}
\caption{Effect of VUV irradiation on the infrared spectrum of grains produced by laser vaporization of an MgSi target in a CO$_2$ atmosphere, with sample substrate holder at 295 to 297~K during irradiation and spectroscopy. Wide view of the spectra (a), focus on the silicate bands (b), on the bands attributed to carbonates (c), and on bands corresponding to CH- and OH-stretching vibrations (d). The raw absorbance spectra, measured consecutively, were offset vertically by an arbitrary amount determined from the average absorbance between 1900 and 2000~cm$^{-1}$ so as to superimpose their baselines.\label{fig:VUV-FTIR-RT-C}}
\end{figure}

Figure~\ref{fig:VUV-FTIR-RT-C} shows the evolution of the carbonated silicate sample as it is irradiated at room temperature with VUV photons. The spectrum measured just before irradiation is slightly different from the spectrum of the fresh deposit (Figure~\ref{fig:MIR-MgSi-CO2}). Specifically, the bands seen between 2800 and 3000~cm$^{-1}$ that correspond to the CH-stretching vibrations of aliphatic hydrocarbon substances grew measurably in the period of several hours between the measurements.

The irradiation procedure attenuates the CH-stretching bands of the aliphatic hydrocarbon substances. We do not observe this phenomenon when irradiating the deposit of silicate grains (Figure~\ref{fig:VUV-FTIR-RT-O}d). We note, though, that the initial aspect of the bands is different because the irradiation of the carbonated sample occured a shorter time after its production (Section~\ref{sec:irrad-Sil}). We propose that the VUV flux desorbs the late-adsorbed aliphatic hydrocarbon contaminants as they would stick to the grains through weak physical adsorption whereas the initially adsorbed aliphatic groups would be chemically bonded to the grain, for instance, as methoxy groups \citep{McDonald58}. Concerning other bands at wavenumbers greater than 2000~cm$^{-1}$, as observed in the experiment with silicate grains, they do not evolve significantly while a weak peak rises at 2233~cm$^{-1}$ and becomes clearly visible after 60~min irradiation. We assign it again to the stretching of Si$\sbond$H bonds \citep{Blanco99} despite the shift from the position observed in the experiment with the silicate deposit (2240~cm$^{-1}$, see Figure~\ref{fig:VUV-FTIR-RT-O}c).

\begin{figure}
\epsscale{1.1}
\plotone{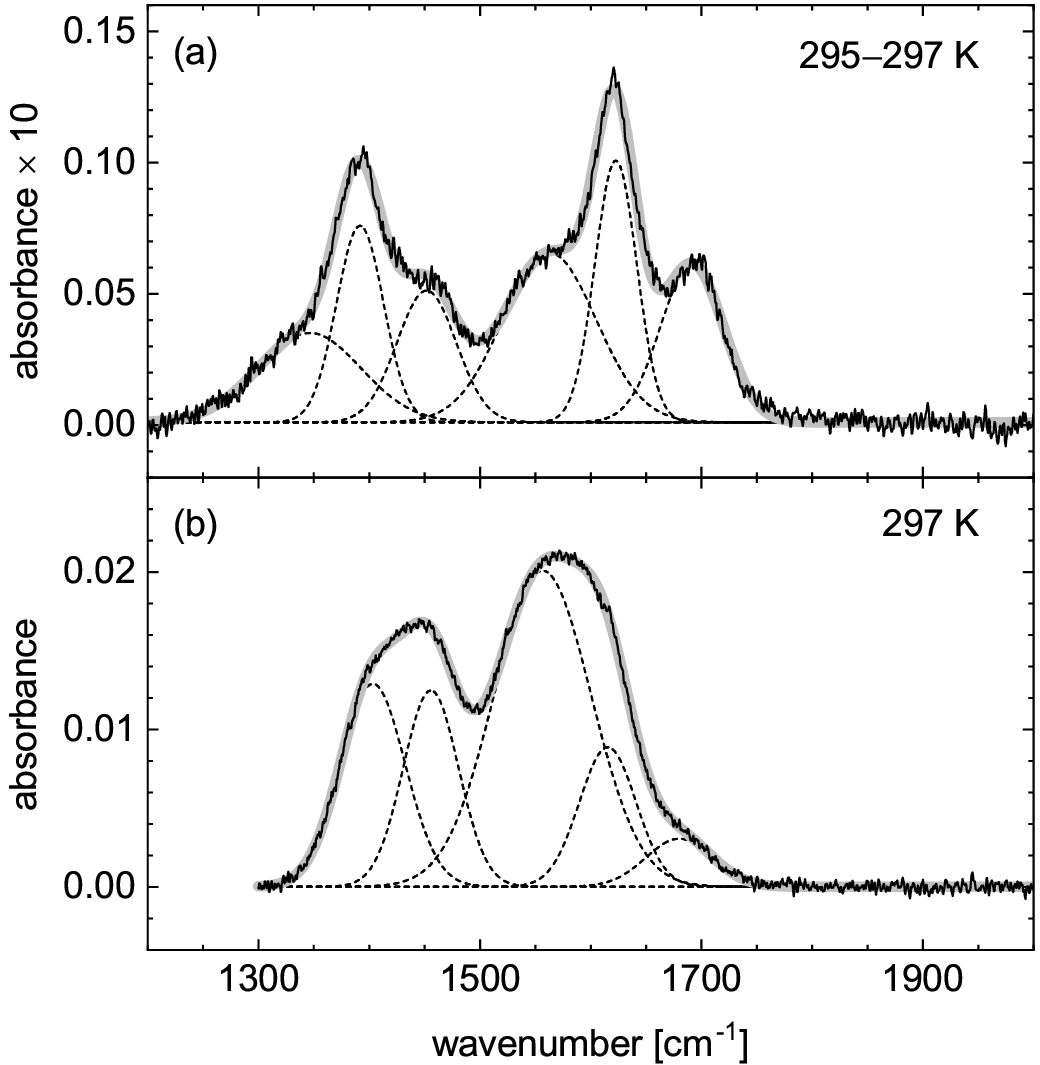}
\caption{Analysis of the absorption features of Figure~\ref{fig:VUV-FTIR-RT-C} attributed to carbonate groups. Difference between the spectra measured before irradiation and after 100~min irradiation (a). Spectrum after 100~min irradiation, after subtraction of a parabola fitted to the local baseline (b). The panels show the measurements (solid curves), fitted Gaussian profiles (dashed curves), and the sum of the fitted profiles (thick gray solid curves).\label{fig:band-analysis-RT}}
\end{figure}

At wavenumbers below 1250~cm$^{-1}$, the peak position of the 10~$\mu$m silicate band gradually shifts from 1043.5 to 1040.4~cm$^{-1}$ as the band widens toward lower wavenumbers, as observed with pure silicate grains (Section~\ref{sec:irrad-Sil}).
As to absorptions attributed to carbonate groups (Section~\ref{sec:MIR-Mix}), we observe in Figure~\ref{fig:VUV-FTIR-RT-C}c that the attenuation of the bands between 1250 and 1800~cm$^{-1}$ tends toward a limit. This indicates that only a fraction of the groups responsible for these absorptions is affected, that is, photodissociated. The difference between the groups that are dissociated and those that are not must lie in their bonding to and position in the silicate structure of the grains. We have analyzed the features in the spectra measured before and after irradiation at room temperature. Figure~\ref{fig:band-analysis-RT}a shows Gaussian profiles fitted to the bands that represent the difference between the two stages, that is, the bands of the chemical structures destroyed upon irradiation. Figure~\ref{fig:band-analysis-RT}b displays profiles fitted to the bands of the structures that remain after irradiation, as the effect of the process was becoming small.

Table~\ref{tbl:bandparam} presents the positions, areas, and FWHM of the Gaussian profiles fitted to the absorption bands in Figure~\ref{fig:band-analysis-RT}. Although the bands common to the spectra show differences in terms of position and width, the differences are not large enough to indicate a change in the chemical nature of the absorbing structures. We conclude that the photodissociated structures and those remaining are of the same nature, namely, carbonate groups, and propose that the latter remain because they are more deeply embedded in the silicate matter. The band at 1346~cm$^{-1}$ is an exception in that respect as it is completely removed by irradiation.

\begin{deluxetable*}{lllllll}
\tablecaption{Band Parameters\tablenotemark{a}\label{tbl:bandparam}}
\tablehead{\colhead{Band} & \multicolumn{3}{c}{Dissociated} & \multicolumn{3}{c}{Remaining} \\ & \colhead{Position} & \colhead{Area} & \colhead{FWHM} & \colhead{Position} & \colhead{Area} & \colhead{FWHM} }
\startdata
1 & 1346 &  0.40 & 109 &  ... &  ... & ... \\
2 & 1392 &  0.40 &  50 & 1403 &  0.94 &  69 \\
3 & 1452 &  0.34 &  63 & 1456 &  0.78 &  59 \\
4 & 1563 &  0.68 &  99 & 1557 &  2.21 & 103 \\
5 & 1623 &  0.47 &  44 & 1615 &  0.57 &  60 \\
6 & 1692 &  0.43 &  67 & 1680 &  0.23 &  71 \\
\enddata
\tablenotetext{a}{In cm$^{-1}$.}
\tablecomments{Values for dissociated and remaining groups correspond to the fitted profiles shown in Figure~\ref{fig:band-analysis-RT}.}
\end{deluxetable*}

\begin{figure}
\epsscale{1.1}
\plotone{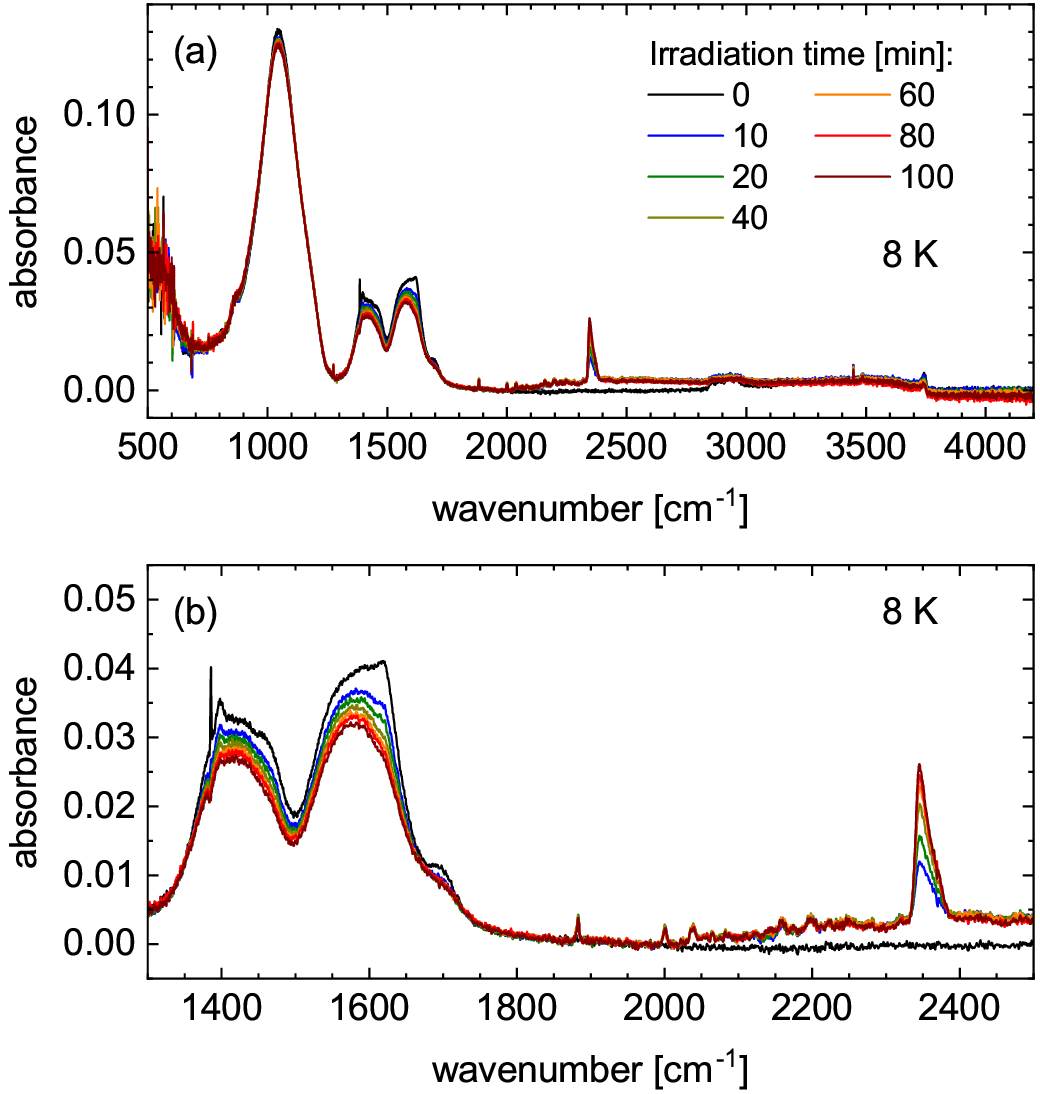}
\caption{Effect of VUV irradiation on the IR spectrum of grains produced by laser vaporization of an MgSi target in a CO$_2$ atmosphere, with sample substrate holder at 8~K during irradiation and spectroscopy. Wide view of the spectra (a) and focus on the carbonate and CO$_2$ bands (b). The raw absorbance spectra, measured consecutively, were offset vertically by an arbitrary amount determined from the average absorbance between 1965 and 1980~cm$^{-1}$ so as to superimpose their baselines.\label{fig:VUV-FTIR-8K}}
\end{figure}

The irradiation experiment was also carried out with a similar sample after the substrate holder was cooled down to 8~K and Figure~\ref{fig:VUV-FTIR-8K} shows the corresponding spectra. The rise of a band peaking at 2345.6~cm$^{-1}$ reveals the formation of CO$_2$ molecules as they accumulate, most likely, on the cold surface of the grains. The absorption previously attributed to the stretching of Si$\sbond$H bonds is not visible whereas an absorption plateau that extends approximately from 2000 to 3200~cm$^{-1}$ and is adorned with numerous minor peaks suggests the presence of a large variety of adsorbed substances following irradiation. After test experiments (see Appendix~\ref{apx:VUV-cont}), we ascribe eventually the plateau and its minor peaks to irradiation-induced F-centers and impurity ions in the KBr substrate that are revealed because of the low temperature \citep{Carlier78,Korovkin93}. Although a test experiment showed adsorbed CO and CO$_2$ molecules on both deposit-covered and deposit-free areas of the substrate, their amount was larger where the deposit laid and they increased in number as the bands attributed to carbonate groups decreased in strength.

\begin{figure}
\epsscale{1.1}
\plotone{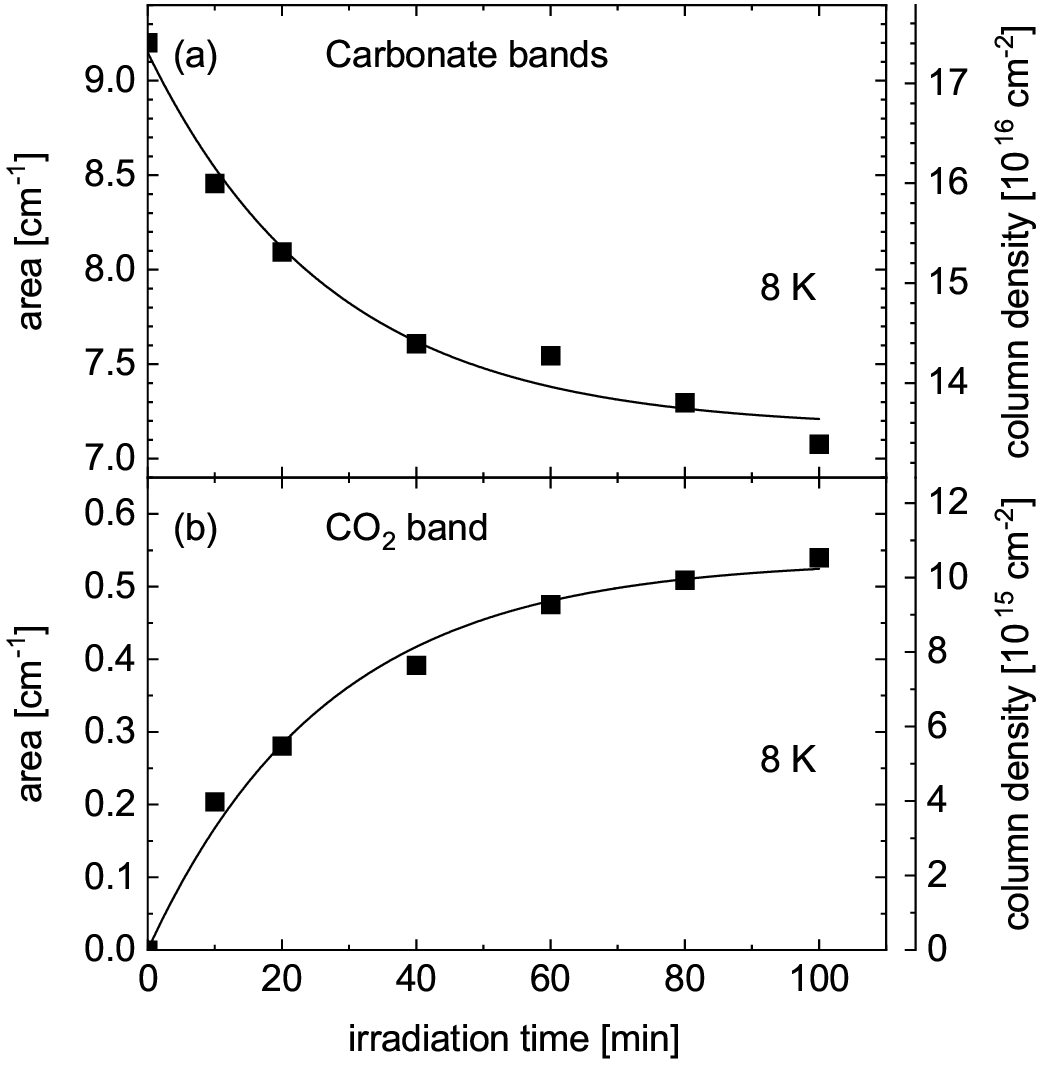}
\caption{Time evolution of band areas in Figure~\ref{fig:VUV-FTIR-8K}. Area of the carbonate bands integrated from 1279 to 1870~cm$^{-1}$ with subtraction of a straight baseline (a). Same for the CO$_2$ band from 2325 to 2389~cm$^{-1}$, with area set to 0 at origin (b). Column density values assume an IR activity or absorption length of 733~km mol$^{-1}$ for the carbonate bands (see Section~\ref{sec:theo}) and a molecular IR activity of 1.18~$\times$ 10$^{-16}$~cm for the CO$_2$ band \citep{Gerakines15} equivalent to 711~km mol$^{-1}$.\label{fig:VUV-FTIR-8K-tau}}
\end{figure}

We observe that the growth rate of the CO$_2$ band in the experiment illustrated with Figure~\ref{fig:VUV-FTIR-8K} is equal to the attenuation rate of the suspected carbonate bands. Figure~\ref{fig:VUV-FTIR-8K-tau} shows the evolution of the band areas and fitted exponential trends with time constants of 27~$\pm$ 2~min and 27~$\pm$ 6~min for CO$_2$ formation and carbonate destruction, respectively. We do not correct the growth rate for the adsorption of CO$_2$ molecules coming from outside the deposit because, when comparing Figures~\ref{fig:VUV-FTIR-8K} and \ref{fig:VUV-FTIR-8K-diff}, we find that the strength of the CO$_2$ band is related to that of the carbonate absorption feature, indicating that the amount of foreign CO$_2$ adsorbed by the grains is negligible in the spectrum of Figure~\ref{fig:VUV-FTIR-8K}. Figure~\ref{fig:VUV-FTIR-8K-tau} also shows the evolution of the bands in terms of column densities. The column density of CO$_2$ molecules increases from nothing to $\sim$10~$\times$ 10$^{15}$~cm$^{-2}$ and that of carbonate groups decreases by $\sim$35~$\times$ 10$^{15}$~cm$^{-2}$. Considering the quality of the IR activities used to derive the column densities, computed from theory for the carbonate groups (this work) and measured with pure ice for the CO$_2$ molecules \citep{Gerakines15}, the quantitative proximity of their opposite variations is remarkable.

\citet{Ciaravella18} proceeded to the irradiation of a synthetic silicate of formula Mg$_2$SiO$_4$ produced with the sol-gel technique that contained carbonate groups present as impurities. The IR spectrum of that material showed bands at 1417 and 1540~cm$^{-1}$, similar to those presently observed. Irradiation of the material cooled to 10~K produced both CO$_2$ and CO molecules. In contrast with their experiment, we do not observe CO molecules in the spectra shown in Figure~\ref{fig:VUV-FTIR-8K}. Yet, in the test performed with another deposit of the same nature and irradiated under the same conditions (see Figure~\ref{fig:VUV-FTIR-8K-diff}c), a clear though weak band revealed the production of CO molecules beside CO$_2$ ones after correcting the spectrum for the contribution of the species responsible for the absorption plateau and minor peaks --irradiation-induced F-centers and impurity ions.

We conclude that VUV irradiation of the deposits dissociates groups that cause two broad absorption bands at $\sim$1420 and $\sim$1580~cm$^{-1}$. The photodissociation produces CO$_2$ and CO molecules, respectively primary and secondary products, consistent with the assignment of the bands to carbonate groups. The effect of VUV irradiation tends toward a limit and we propose that the photostability of the groups depends on their position in the silicate matrix of the grains, at the surface or below.  Furthermore, the time constant of 27~$\pm$ 6~min for a Lyman-$\alpha$ photon irradiance on the order of (145~$\pm$ 50)~$\times$ 10$^{12}$~cm$^{-2}$ s$^{-1}$ (Section~\ref{sec:exp}) means that the carbonate groups that can be photodissociated have a photodissociation cross section on the order of (0.04~$\pm$ 0.02)~$\times$ 10$^{-16}$~cm$^2$.

\subsection{Theoretical Calculations}\label{sec:theo}

\begin{figure}
\epsscale{1.1}
\plotone{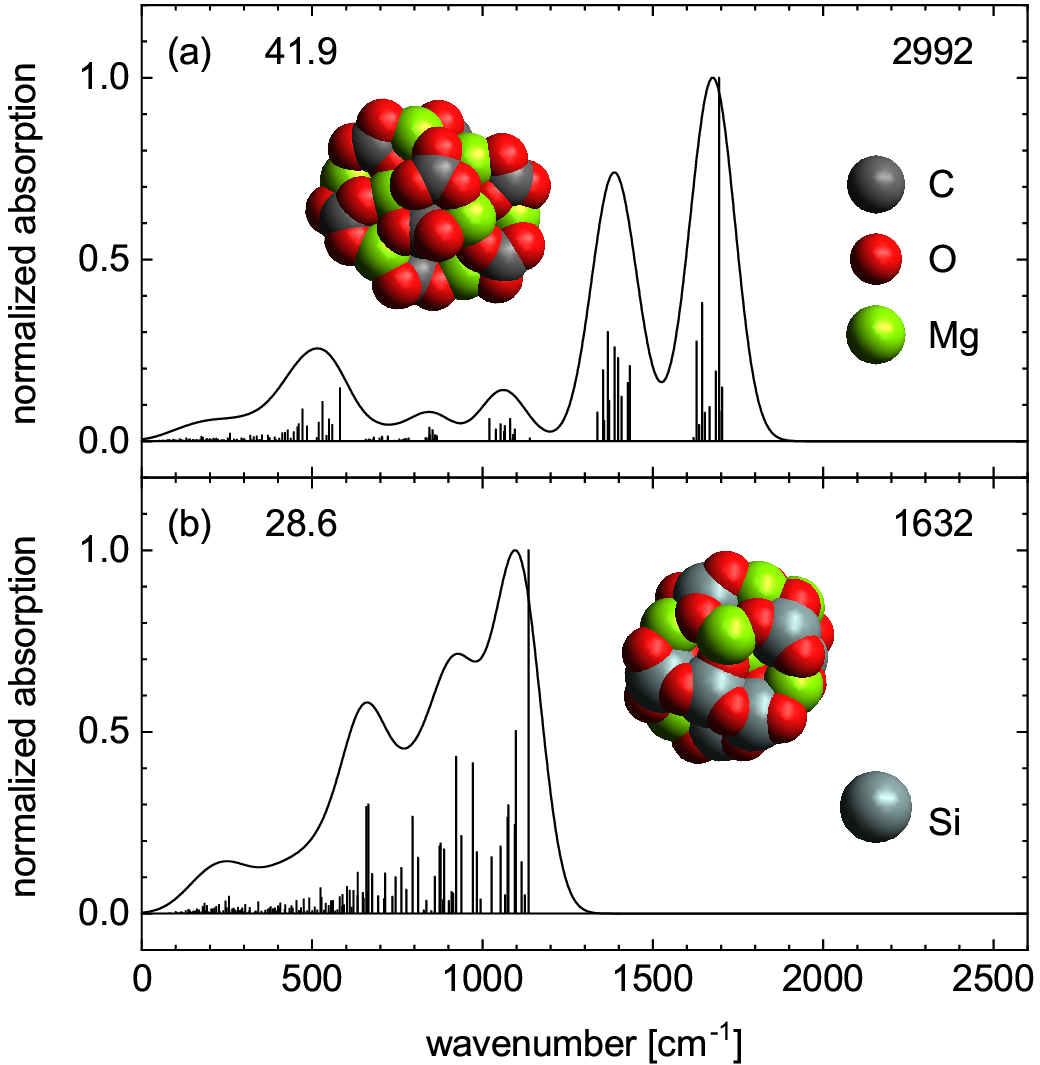}
\caption{Theoretical IR activity of an (MgCO$_3$)$_{10}$ carbonate cluster (a) and an (MgSiO$_3$)$_{10}$ silicate cluster (b). Sticks represent the relative intensities of the fundamental vibrational modes as a function of their unscaled harmonic frequencies. Convolution of the sticks with a Gaussian profile gives the curves. The numbers displayed in each panel are IR intensities in km mol$^{-1}$: maximum of the convoluted spectrum (left) and maximum of the stick spectrum (right). Artworks show the amorphous clusters to which corresponds the stick spectra of both panels, with C, O, Mg, and Si atoms in dark gray, red, green, and light gray colors.\label{fig:theoIR1}}
\end{figure}

Figure~\ref{fig:theoIR1} illustrates the IR activity of vibrational modes for theoretical carbonate and silicate clusters with respective formulas (MgCO$_3$)$_{10}$ and (MgSiO$_3$)$_{10}$. Parentheses in the formulas are employed only to underscore the stoichiometry of the composition and do not indicate any polymeric character. \citet{Chen15} and \citet{Escatllar19} calculated the original structures of the carbonate and silicate clusters, respectively. We have re-optimized them and calculated their vibrational modes at the same B3LYP/6-31g(d) level of theory for consistency (Appendix~\ref{sec:calc}). Figure~\ref{fig:theoIR1} comprises stick spectra that report the IR activity of the fundamental vibrational modes as a function of their unscaled harmonic frequencies. It includes synthetic absorption spectra obtained by convoluting the stick spectra with a Gaussian profile characterized by an FWHM of 140~cm$^{-1}$ so as to obtain spectra that resemble those of the grains produced by gas-phase condensation. The area of each synthetic spectrum is equal to the total IR activity of the vibrational modes. We do not take into account absorption from thermally populated vibrational levels, for the measured absorption spectrum of carbonated silicate grains is the same at 295~K (Figure~\ref{fig:MIR-MgSi-CO2}) and 10~K (not shown).

The synthetic spectrum of the theoretical carbonate grain shows a series of peaks at 516, 843, 1061, 1387, and 1676~cm$^{-1}$. The first one corresponds to the stretching of Mg$\sbond$O bonds, and the others to vibrational modes of the CO$_3$ carbonate groups. They are the out-of-plane bending or inversion mode ($\nu_2$, 843~cm$^{-1}$), their symmetric stretching mode ($\nu_1$, 1061~cm$^{-1}$) activated by the asymmetry of the environment, and the degeneracy-lifted asymmetric stretching mode ($\nu_3$, 1387 and 1676~cm$^{-1}$). Absorption related to the in-plane bending mode of carbonate groups ($\nu_4$) does not give a distinct band because the IR activity of the mode is low (less than 40~km~mol$^{-1}$, see the sticks between 650 to 800~cm$^{-1}$ in Figure~\ref{fig:theoIR1}a) and the deformation various (see the spread of the sticks) because of the amorphous nature of the structure.

The spectrum of the theoretical silicate cluster is dominated by a single peak at 1096~cm$^{-1}$ that corresponds to the asymmetric stretching of Si$\sbond$O bonds. The multiple vibrational modes that underlie the synthetic bands attributed to the deformation of SiO$_4$ tetrahedra in an amorphous structure do not exclusively involve specific types of tetrahedra, where the number of non-bridging O atoms defines the type. Out of the ten SiO$_4$ tetrahedra found in this specific cluster, two, six, and two feature one, two, and three non-bridging O atoms, respectively. The small size of the cluster likely prevents the inclusion of a tetrahedra with zero or four non-bridging O atoms.

\begin{figure}
\epsscale{1.1}
\plotone{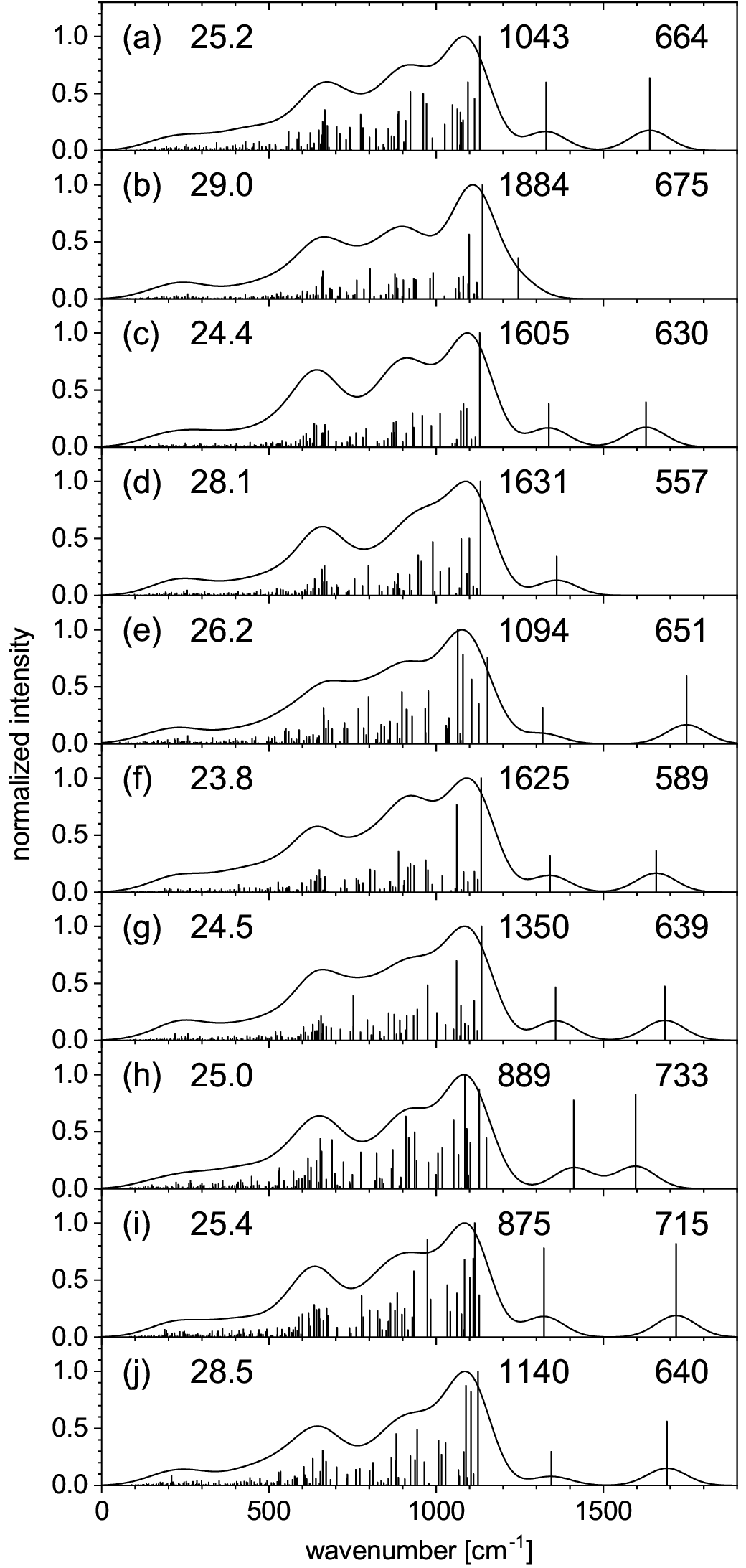}
\caption{Theoretical IR activity of ten (MgCO$_3$)(MgSiO$_3$)$_{9}$ carbonated silicate clusters derived from a same (MgSiO$_3$)$_{10}$ enstatite-like structure (a to g). Sticks represent the relative intensities of the fundamental vibrational modes as a function of their unscaled harmonic frequencies. The convolution of the sticks with a Gaussian profile gives the curves. The numbers displayed in each panel are IR intensities in km mol$^{-1}$: from left to right, maximum of the convoluted spectrum, maximum of the stick spectrum, and value for the strongest CO-stretching stick.\label{fig:theoIR2}}
\end{figure}

Figure~\ref{fig:theoIR2} presents stick spectra of the IR-active vibrational modes of carbonated enstatite-like clusters with the formula (MgCO$_3$)(MgSiO$_3$)$_{9}$. Stick positions are the unscaled harmonic frequencies of the modes. We derived the structures from that of a unique (MgSiO$_3$)$_{10}$ silicate cluster (Appendix~\ref{sec:calc}), the spectrum of which is shown in Figure~\ref{fig:theoIR1}. The vibrational modes with wavenumbers greater than 1200~cm$^{-1}$ are exclusively asymmetric CO-stretching modes of the CO$_3$ group and none of the modes with a wavenumber lower than 1200~cm$^{-1}$ corresponds to asymmetric CO stretching.

Quantum chemistry calculations have given us a picture of the nanometer-sized grains in terms of electric charge distribution and natural bond orbitals (NBOs, not to be confused with non-bridging O atoms). Various assemblages of C and O atoms appear among the ten theoretical clusters. Two of them show a C atom covalently bonded to four O atoms instead of three, thus giving the spectra of Figures~\ref{fig:theoIR2}b and \ref{fig:theoIR2}d. Moreover, the CO$_n$ ($n$ = 3 or 4) units are in general not pure ions as their O atoms are often covalently bonded with Si atoms as well. When they are not, Coulomb interaction attaches them to neighboring Mg or Si atoms, for electric charges in these structures are on the order of -1, 1.3, and 2~e for O, Mg, and Si atoms, respectively. The diversity of the bonds between a CO$_3$ unit and its environment lifts the degeneracy of the asymmetric CO-stretching modes as observed in the measured IR spectra.

We remark that the maximum IR activity of the SiO-stretching modes in the spectra of Figure~\ref{fig:theoIR2} varies from 875 to 1884~km mol$^{-1}$. The position and intensity of the SiO-stretching modes that arise in a single amorphous grain vary greatly, though. Consequently, the maximum IR activity of the convoluted spectra in Figure~\ref{fig:theoIR2} is similar for all grains, ranging from 23.8 to 29.0~km mol$^{-1}$ and averaging at 26.0~km mol$^{-1}$. We note that the corresponding value derived for the parent silicate structure (Figure~\ref{fig:theoIR1}b) is 28.6~km mol$^{-1}$ and lies in this interval. Thus carbonate groups dispersed in an amorphous silicate matrix do not affect the silicate spectrum strongly when their concentration is as high as $\sim$10\%. Still in Figure~\ref{fig:theoIR2}, the IR activity of the CO-stretching modes vary little among the stick spectra, the strongest mode in each spectrum giving a value ranging from 557 to 733~km mol$^{-1}$. As these modes are well separated from the others in terms of position, the observation applies to the convoluted spectra too. Interestingly, the maximum strength of such a mode can be as high as 2992~km mol$^{-1}$ in a pure carbonate cluster (Figure~\ref{fig:theoIR1}a). The cause is the collective involvement of multiple carbonate groups in a single stretching mode despite the amorphous structure of the cluster.

Only the theoretical spectrum of Figure~\ref{fig:theoIR2}h exhibits a separation of the two asymmetric CO-stretching modes close to that observed in the experiments, that is, $\sim$200~cm$^{-1}$. Taking into account the fact that the observed absorptions comprise multiple components does not affect the comparison. The computed positions of the bands are then 1412 and 1596~cm$^{-1}$, similar to the measured positions of the absorptions that we have attributed to carbonate groups, $\sim$1420 and $\sim$1580~cm$^{-1}$ (see Figures~\ref{fig:MIR-MgSi-CO2}, \ref{fig:MIR-MgSi-O2CO2}, and \ref{fig:VUV-FTIR-RT-C}--\ref{fig:VUV-FTIR-8K}). In contrast, the separation in the other theoretical spectra ranges from 290~cm$^{-1}$ (Figure~\ref{fig:theoIR2}c) to 430~cm$^{-1}$ (Figure~\ref{fig:theoIR2}e) and is thus in every case too large to match the measured value, even considering a correction of the vibrational frequency scale that generally amounts to a few percent. The individual theoretical structures are, however, results of an arbitrary construction method (Appendix~\ref{sec:calc}). Yet, one of them produces an IR spectrum that coincides with those measured in the experiments. That it is the structure that gives the smallest splitting of the degenerated $\nu_3$ mode of the carbonate group may not be a coincidence. A theoretical approach that takes into account the last stage of grain condensation, namely, cooling, would give more relevant structures and spectra, and new information to evaluate the assignment of the experimental spectra.

\begin{figure}
\epsscale{1.1}
\plotone{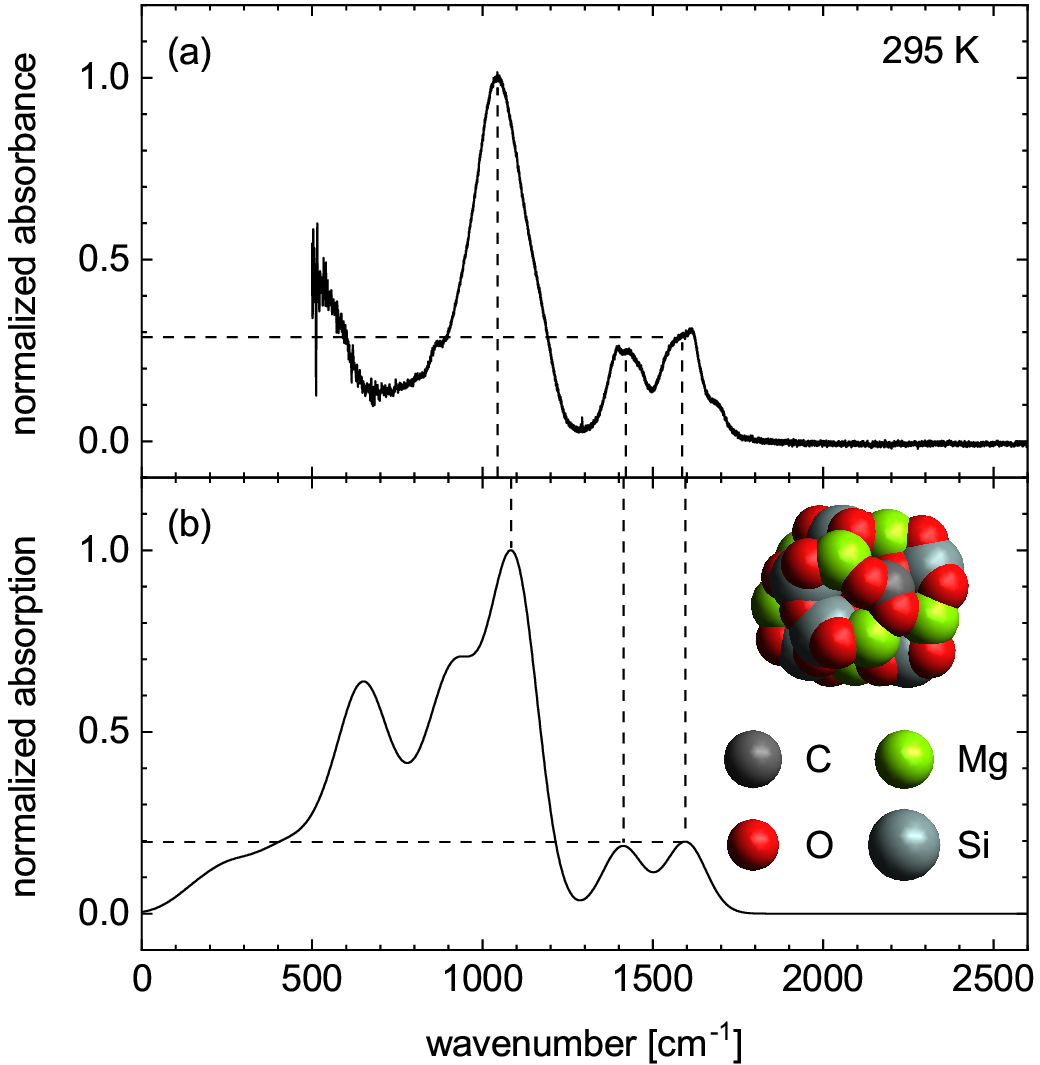}
\caption{Comparison of measured and theoretical IR spectra. (a) Spectrum of carbonated amorphous MgSiO$_3$ silicate grains (from Figure~\ref{fig:MIR-MgSi-CO2}). (b) Theoretical IR activity of an (MgCO$_3$)(MgSiO$_3$)$_{9}$ carbonated silicate cluster (same as Figure~\ref{fig:theoIR2}h). Dashed lines are guides to the eye.\label{fig:expvstheo}}
\end{figure}

Finally, Figure~\ref{fig:theoIR2}h shows that the peak intensity of a CO-stretching band represents $\sim$20\% of that of the SiO-stretching band when there is one C atom for nine Si atoms in the grains. For an average peak activity of 26.0~km mol$^{-1}$ computed for the SiO-stretching band from the values given in Figure~\ref{fig:theoIR2}, the activity of the CO-stretching band arising from a single carbonate group is then 5.2~km mol$^{-1}$. Thus, concerning asymmetric stretching modes, the IR activity of a carbonate group is in average almost twice as large as that of a silicate group in the case of amorphous carbonated silicate grains. Additionally, comparison of the relative SiO- and CO-stretching band peak intensities in the theoretical and measured spectra in Figure~\ref{fig:expvstheo} suggests that the grains produced by gas-phase condensation contain between one and two C atoms for nine Si atoms.

\section{Discussion}

The presence of carbonates in cosmic dust was first mentioned by \citet{Gillett73}. They observed that an emission feature at 11.3~$\mu$m in the spectra of two PNe, namely, NGC~7027 and BD~+30$^\circ$3639, showed similarities to an absorption band of carbonates, which would correspond to the $\nu_2$ out-of-plane deformation mode of CO$_3^{2-}$ groups. Assuming the assignment to carbonates was correct, \citet{Bregman75} proposed that they consisted, in NGC~7027, of MgCO$_3$ impurities in dust grains so as to fit the observed wavelength and explain the resilience of the substance.

Another emission band was used to identify the presence of carbonates in the PNe NGC~6302 and NGC~6537 \citep{Kemper02a}. An emission band at 92.6~$\mu$m, as measured in the spectrum of NGC~6302, could only be attributed to the lattice vibration of calcium carbonate, specifically calcite \citep{Kemper02a,Kemper02b}. Though the spectrum of NGC~6302 did not reveal any emission band at 7~$\mu$m \citep{Molster01}, that is, the wavelength of the strongest IR-active vibration for carbonates \citep[e.g.,][]{Dorschner80}, the low temperature of the dust would have definitely prevented its detection. Indeed, the analysis of the spectrum yielded a temperature of 30--60~K considering the various components of the dust \citep{Kemper02a}.

The attribution of the 11.3~$\mu$m and 92.6~$\mu$m emission bands to carbonates triggered discussions. Concerning the emission at 11.3~$\mu$m, \citet{Russell77b} could not explain the lack of a companion at 7~$\mu$m in the observations of NGC~7027, even by taking temperature into account. Then, after a detailed discussion that included quantitative elements, \citet{McCarthy78} concluded that the assignment of the 11.3~$\mu$m to ordinary crystalline carbonates was uncertain. As to the assignment of the 92.6~$\mu$m emission in NGC~6302 to calcite \citep{Kemper02a,Kemper02b}, \citet{Ferrarotti05} remarked that the observed material had to have formed in stellar winds \citep[see also][]{Cohen74} and that modeling failed to explain calcite formation in the corresponding conditions. Nonetheless, the experiments by \citet{Toppani05} suggested a route compatible with the conditions.

\begin{figure}
\epsscale{1.1}
\plotone{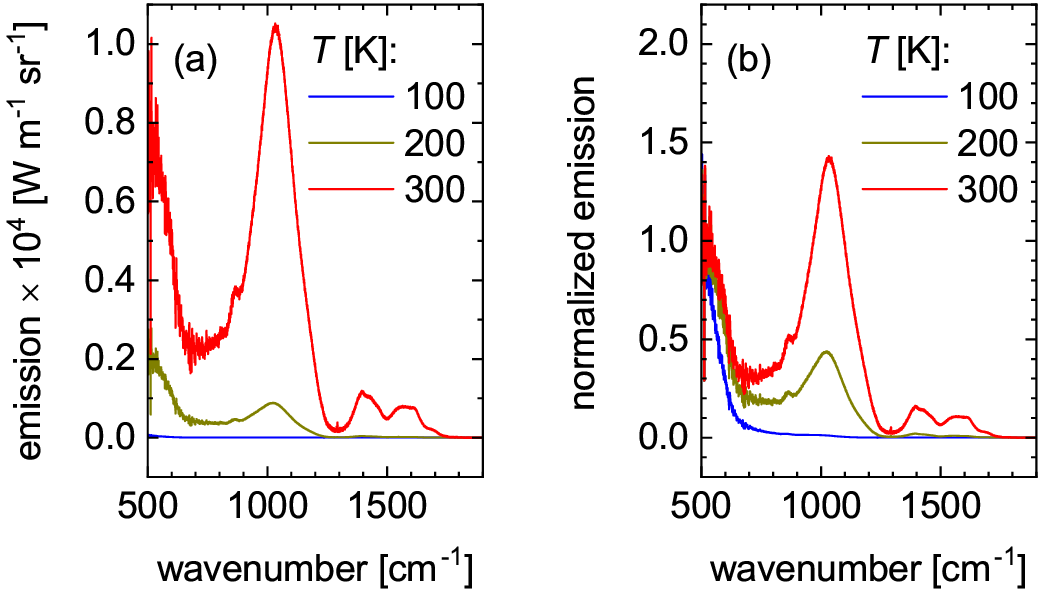}
\caption{Simulated emission spectra of amorphous carbonated magnesium silicate grains at three temperatures. Absolute spectra (a) and spectra normalized at the fitted position of the 20~$\mu$m band (b). Each curve corresponds to the measured spectrum of Figure~\ref{fig:MIR-MgSi-CO2} multiplied by a Planck function in wavenumber representation at the chosen temperature $T$.\label{fig:IRemission}}
\end{figure}

The studies and issues mentioned above considered crystalline materials such as calcite while we are looking at carbonate groups embedded in amorphous magnesium silicates. While lattice vibration bands such as the band of calcite near 92~$\mu$m relate only to crystalline grains, bands characteristic of the carbonate groups such as the 7~$\mu$m band arise independently of the grain structure and can reveal carbonated amorphous silicates. The condensation of silicate grains occurs in outflows of evolved oxygen-rich stars \citep[e.g.,][]{Hoefner09}. In these outflows, according to modeling and observations \citep[][and references therein]{Gobrecht16}, the C atoms are contained in CO molecules and in much less abundant species that include CO$_2$. For comparison, CO$_2$ is four to five orders of magnitude less abundant than CO and two to three orders of magnitude less abundant than SiO. \citet{Gobrecht16} argued that CO does not contribute to the formation of dust clusters, precursors of grains. As to CO$_2$, its very low abundance makes it a minor reactant in a potential dust formation process. Still, we note that, as both CO and H$_2$O molecules are abundant in the regions of silicate condensation, the carbonates observed in PNe may have formed from just-formed silicate grains in a process that involves these species as proposed by \citet{Toppani05}. Certainly, spectra of PNe do not feature, up to now, any band in the 7~$\mu$m region that we can doubtlessly attribute to carbonate groups, indicating the low carbonate content of silicates in warm dust. Bands near this wavelength would not be visible in emission spectra of cold dust as illustrated with Figure~\ref{fig:IRemission}.

Concerning protostars, their outflows are not typical sites of grain condensation. Moreover, the central star irradiates them with energetic photons, including Lyman-$\alpha$ ones \citep[][and references therein]{Bally16} that would dissociate carbonate groups if they were to form on the surface of grains. The carbonates observed in protostars have rather formed by processing of ices on grains of the surrounding nebular dust \citep{Ceccarelli02,Chiavassa05}. In relation to this hypothesis, we observed the formation of carbonates in a study on the condensation of refractory materials at cryogenics temperatures \citep{Rouille20}. The condensation proceeded as atoms and molecules considered to be precursors of silicates and solid carbon diffused in Ne ice. Beside these precursors, which consisted in particular of Mg and Fe atoms, SiO and C$_n$ ($n$ = 2--10) molecules, CO, CO$_2$ and H$_2$O molecules were also present in large numbers. The result suggests that ice chemistry in the nebular cloud may have produced the carbonates observed in protostars. When ices are present, however, the complexity of absorption spectra in the 5.5--7.5~$\mu$m range \citep{Boogert08,Boogert15} strongly hinders the detection of the CO-stretching bands of carbonate groups.

Finally, the carbonate content of cosmic dust currently represents, to our knowledge, an indeterminate quantity of oxygen atoms that are depleted from the interstellar gas phase. That this quantity could solve the issue of the missing interstellar O atoms introduced by \citet{Jenkins09,Jenkins19} and \citet{Whittet10} is unlikely. On the one hand, about 50\% of the expected O atoms are unaccounted for in the densest interstellar regions, while 20\% are locked in silicates and oxides, and the remaining in ices and in the gas phase \citep{Whittet10}. On the other hand, considering (i) the clear observation of interstellar silicates by absorption at 10~$\mu$m, (ii) the absence of an indisputable detection of carbonates by absorption at 7~$\mu$m in the same regions, and (iii) the similar oscillator strengths of the two vibrational modes involved in these absorptions, we find that the abundance of interstellar carbonates or carbonate groups is much lower than that of silicates. Thus, only a very small amount of oxygen is locked in carbonates and additional materials need to be evaluated as oxygen reservoirs, for instance, organic refractory materials \citep{Whittet10,Jones19}.

\section{Conclusion}

Amorphous carbonated magnesium silicate grains form through the non-TE condensation of atoms following the pulsed laser ablation of an MgSi target in a CO$_2$ atmosphere. The atoms condense simultaneously, for the grains consist of an amorphous MgSiO$_3$ silicate matrix through which dispersed CO$_3$ carbonate groups are homogeneously distributed. Accordingly, the IR spectra of the grains feature the characteristic main bands of amorphous silicates at $\sim$10 and $\sim$20~$\mu$m and two bands at $\sim$6.3 and $\sim$7.0~$\mu$m that we attribute to the degeneracy-lifted asymmetric CO stretching mode of the carbonate groups. Despite an estimated Si:C ratio of 9:1 to 9:2, the dispersion of carbonate groups in the amorphous magnesium silicate matrix does not alter significantly the profile of the 10~$\mu$m band. Consequently, measurements of this major band in astronomical spectra may not reveal whether the grains are carbonated and possibly contain up to 11\%--18\% more oxygen than expected. In absorption spectra, the identification of amorphous carbonated magnesium silicate grains can rely on the detection of the doublet at $\sim$6.3 and $\sim$7.0~$\mu$m. Regarding emission observations, the relatively long wavelength of the double band and the absence of lattice vibration band make it impractical to detect amorphous carbonated silicate grains in cold regions.

We find that the irradiation of the amorphous carbonated magnesium silicate grains with Lyman-$\alpha$ photons dissociates the carbonate groups at the surface of the grains and produces CO$_2$ molecules. The corresponding photodissociation cross section is (0.04~$\pm$ 0.02)~$\times$ 10$^{-16}$~cm$^2$.

The experimental conditions in which the carbonated silicate grains form are relevant to those that prevail in winds and outflows of stellar objects. The formation of cosmic silicate grains, however, occurs in the outflows of oxygen-rich evolved stars where carbon is mainly locked in CO molecules that are considered to not play a role in dust formation. The carbonates observed in PNe may have formed in outflows from just-formed silicate grains in a process that features CO and H$_2$O molecules, both being relatively abundant. Concerning the outflows of protostars, they do not normally produce silicate dust and energetic photons that irradiate them would dissociate carbonate groups. The carbonates detected in protostars must be the products of ice chemistry.

\acknowledgments
The authors are thankful to the Deutsche Forschungsgemeinschaft (DFG) for its support through project No. 451244650.

\bibliographystyle{aasjournal}

\appendix

\restartappendixnumbering

\section{Effect of Ambient Gas Composition}\label{apx:gas-comp}

\begin{figure}
\epsscale{1.1}
\plotone{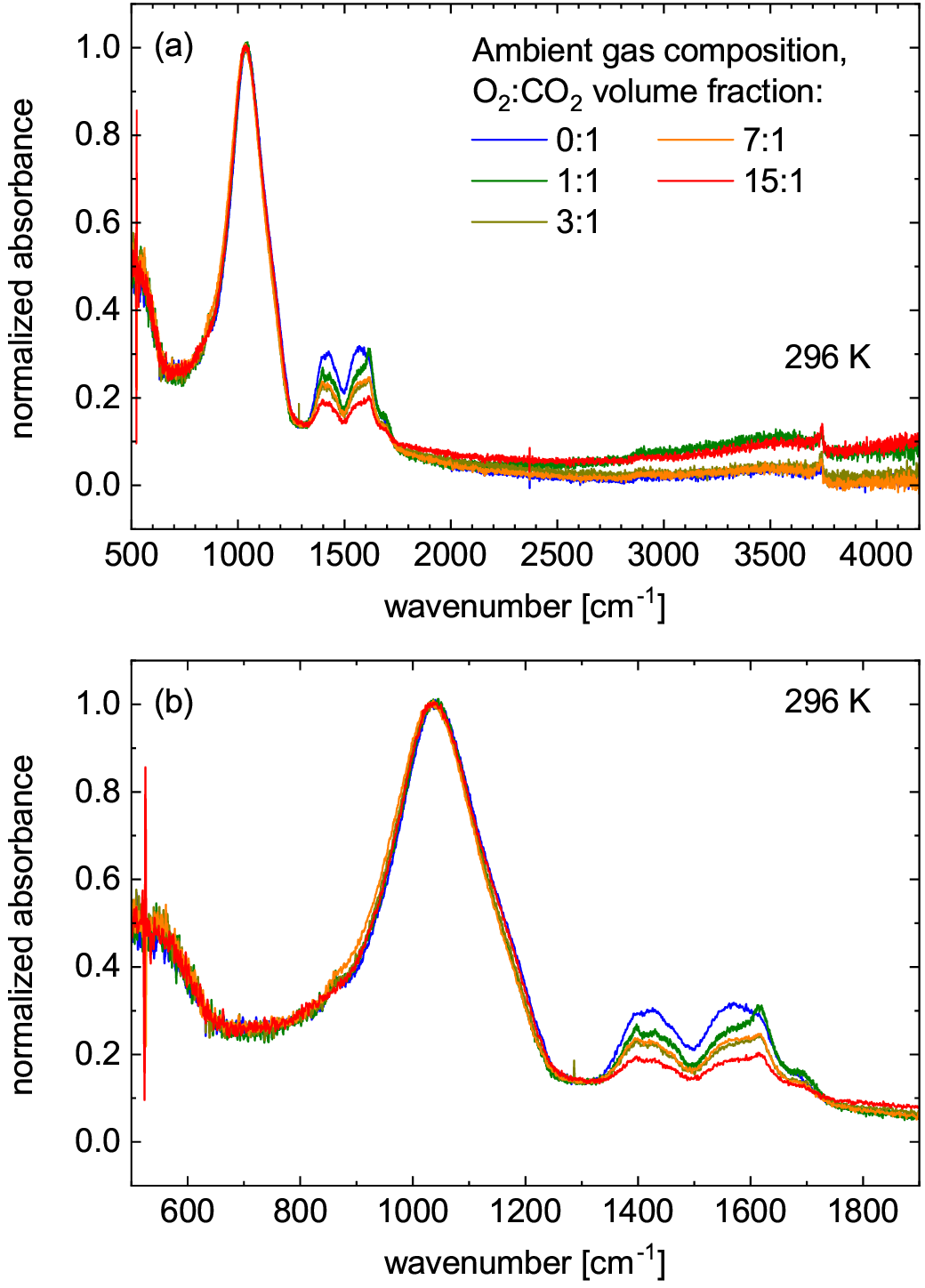}
\caption{Absorption spectra of grains obtained from laser vaporization of an MgSi target in O$_2$:CO$_2$ gas mixtures. Broad range measurements carried out in situ at 296~K (a) and focus on the main bands (b). The average absorbance between 1305 and 1315~cm$^{-1}$ serves as reference for offset and the maximum of the 10~$\mu$m band as reference for normalization. The baselines are not further modified.\label{fig:MIR-MgSi-O2CO2}}
\end{figure}

The relative strengths of the bands assigned to silicate and carbonate groups depend on the composition of the ambient gas used during laser ablation. Figure~\ref{fig:MIR-MgSi-O2CO2} displays spectra of carbonated silicate grains synthesized with different O$_2$:CO$_2$ mixtures and they show that the bands of carbonate groups gain in relative strength as the concentration of CO$_2$ increases. The trend is regular although the spectra measured in the experiments with the 3:1 and 7:1 gas mixtures are similar. Absorption bands caused by CH- and OH-stretching vibrations do not evolve similarly, in accordance with the assignment concerning carbonate groups.

\section{Analysis of Spectra}\label{apx:MIR-fits}

\begin{figure}
\epsscale{1.1}
\plotone{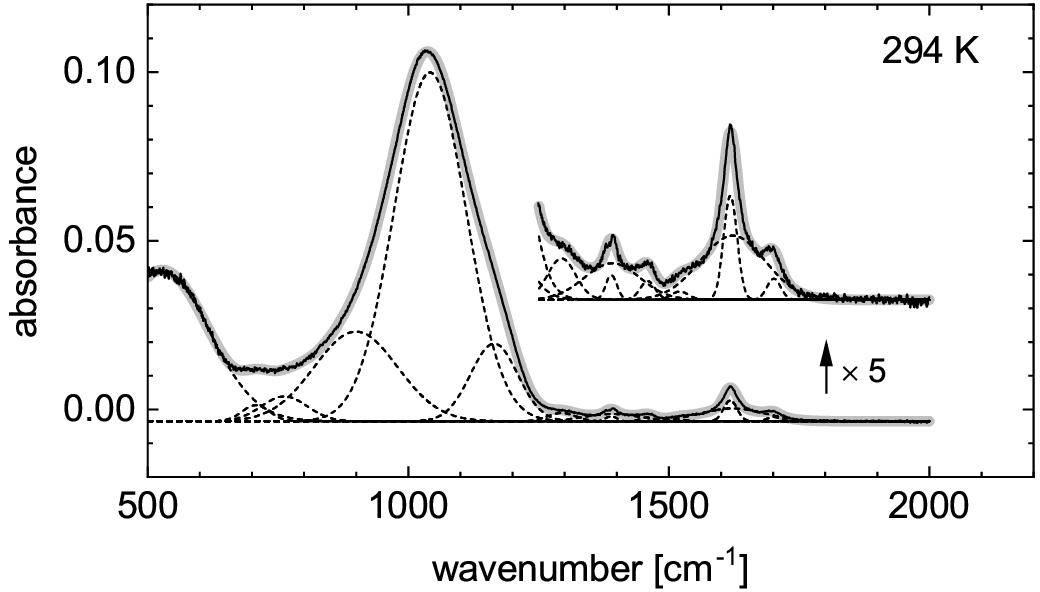}
\caption{Analysis with Gaussian profiles of the IR spectrum of silicate grains. The 1250--2000~cm$^{-1}$ range is reproduced vertically shifted and expanded for a better display of details. The curves correspond to the measured spectrum (black solid curve), fitted Gaussian profiles (black dashed curves), and the sum of the fitted profiles (thick gray solid curve).\label{fig:MIR-MgSi-O2-fits}}
\end{figure}

Figure~\ref{fig:MIR-MgSi-O2-fits} shows the decomposition of the spectrum displayed in Figure~\ref{fig:MIR-MgSi-O2}. The calculated spectrum fitted to the measurements comprises 14 Gaussian peaks described in Table~\ref{tbl:bandparam2}. Narrow and broad components are superimposed at 1390 and $\sim$1620~cm$^{-1}$.

\begin{deluxetable}{llll}
\tablecaption{Band Parameters\tablenotemark{a}\label{tbl:bandparam2}}
\tablehead{\colhead{Band} & \colhead{Position} & \colhead{Area} & \colhead{FWHM} }
\startdata
1  &  528 & 10.01 & 210 \\
2  &  707 &  0.31 &  62 \\
3  &  762 &  0.81 & 103 \\
4  &  899 &  5.32 & 188 \\
5  & 1042 & 17.88 & 163 \\
6  & 1165 &  2.60 & 106 \\
7  & 1294 &  0.16 &  63 \\
8  & 1390 &  0.04 &  25 \\
9  & 1391 &  0.29 & 125 \\
10 & 1459 &  0.04 &  34 \\
11 & 1521 &  0.02 &  38 \\
12 & 1618 &  0.20 &  30 \\
13 & 1623 &  0.59 & 147 \\
14 & 1702 &  0.04 &  31 \\
\enddata
\tablenotetext{a}{In cm$^{-1}$.}
\tablecomments{The values correspond to the fitted profiles shown in Figure~\ref{fig:MIR-MgSi-O2-fits}.}
\end{deluxetable}

\section{Irradiation-induced Contamination}\label{apx:VUV-cont}

\begin{figure}
\epsscale{1.1}
\plotone{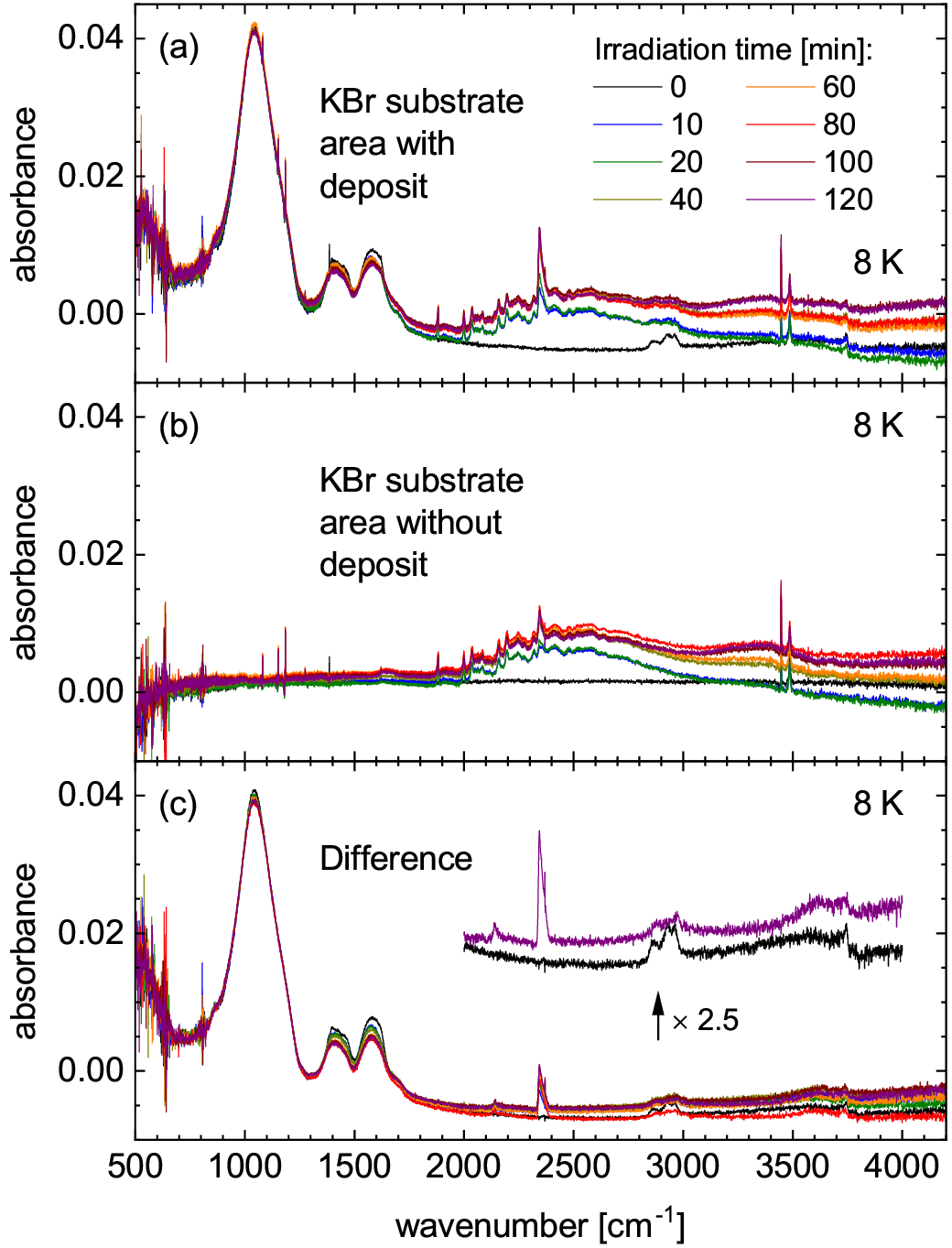}
\caption{Effect of VUV irradiation on the IR spectrum of grains produced by laser vaporization of an MgSi target in a CO$_2$ atmosphere, with sample substrate holder at 7--8~K during irradiation and spectroscopy. Spectra measured at areas of the substrate with deposit (a) and without deposit (b). Difference between spectra with and without deposit (c). The 2000--4000~cm$^{-1}$ range of the spectra measured before irradiation and after 120~min irradiation are reproduced vertically shifted and expanded for easier comparison.\label{fig:VUV-FTIR-8K-diff}}
\end{figure}

We carried out test experiments to determine the origin of the species appearing during VUV irradiation at 8~K (see Figure~\ref{fig:VUV-FTIR-8K}). First, we measured the spectra (not shown) of a bare KBr substrate irradiated while under vacuum and cold (7--8~K) and observed the appearance of an absorption plateau with minor peaks, the narrow bands of CO and CO$_2$ at 2137--2138 and 2341--2342~cm$^{-1}$, respectively, and water bands. The molecules came obviously from components of the vacuum chamber and cryocooler. Second, as illustrated with Figure~\ref{fig:VUV-FTIR-8K-diff}, we compared the spectra of two areas of a single substrate, one bare and the other covered with carbonated silicate grains, irradiated simultaneously while under vacuum and at low temperature. We found the same amounts of new species when probing both areas. We inferred from these tests that the species that caused the plateau and the minor peaks of proportional strength were either emanating from the vacuum chamber or being created in the KBr substrate. While the small amount of CO$_2$ found on the deposit-free area likely originated in the chamber, CO being a secondary photodissociation product, we attributed the plateau and the minor peaks to F-centers and impurity ions induced in KBr by irradiation \citep{Carlier78,Korovkin93}. The lifetime of the centers is long at low temperature.

We remark that the contribution of the KBr substrate to the spectra in Figure~\ref{fig:VUV-FTIR-8K-diff} does not depend on the presence of the grain deposit. With a thickness of 97~nm (assuming a density of 2.71 g cm$^{-3}$ and not taking porosity into account), the deposit does not attenuate measurably the VUV radiation that creates the F-centers and impurity ions in the substrate. Attenuation is visible when we compare Figures~\ref{fig:VUV-FTIR-8K-diff} and \ref{fig:VUV-FTIR-8K}, however. They show that the height of the absorption plateau varies inversely to that of the 10~$\mu$m silicate band.

\end{document}